\documentclass[fleqn,usenatbib]{mnras}

\usepackage[T1]{fontenc}
\usepackage{ae,aecompl}
 
\usepackage{graphicx}	% Including figure files
\usepackage{amsmath}	% Advanced maths commands

\usepackage{amssymb}	% Extra maths symbols

\title[Intrinsic absorber in IGR~J17091-3624]{Simultaneous detection of an intrinsic absorber and a compact jet emission in the X-ray binary IGR J17091-3624 during a hard accretion state} 

\author[Gatuzz et al.]{
E. Gatuzz$^{1}$\thanks{E-mail: efraingatuzz@gmail.com},
M. D\'iaz Trigo$^{1}$,
J.C.A. Miller-Jones$^{2}$
and S. Migliari$^{3,4}$
\\
$^{1}$ESO, Karl-Schwarzschild-Strasse 2, D-85748 Garching bei M\"unchen, Germany\\ 
$^{2}$International Centre for Radio Astronomy Research, Curtin University, G.P.O. Box U1987, Perth, WA, 6845, Australia\\
$^{3}$XMM-Newton Science Operations Centre, ESAC/ESA, Camino Bajo del Castillo s/n, Urb. Villafranca del Castillo, \\
28691 Villanueva de la Ca\~nada, Madrid, Spain\\
 $^{4}$Institute of Cosmos Sciences, University of Barcelona, Mart\'i i Franqu\`es 1, 08028 Barcelona, Spain\\
} 
\date{Accepted XXX. Received YYY; in original form ZZZ} 
\pubyear{2018} 
\begin{document}
 \label{firstpage}
\pagerange{\pageref{firstpage}--\pageref{lastpage}}
\maketitle 
\begin{abstract}
We present a detailed analysis of three {\it XMM-Newton} observations of the black hole low-mass X-ray binary IGR~J17091-3624 taken during its 2016 outburst. Radio observations obtained with the Australia Telescope Compact Array (ATCA) indicate the presence of a compact jet during all observations. From the best X-ray data fit results we concluded that the observations were taken during a transition from a hard accretion state to a hard-intermediate accretion state. For Observations~1 and 2 a local absorber can be identified in the EPIC-pn spectra but not in the RGS spectra, preventing us from distinguishing between absorption local to the source and that from the hot ISM component. For Observation~3, on the other hand, we have identified an intrinsic ionized static absorber in both EPIC-pn and RGS spectra. The absorber, observed simultaneously with a compact jet emission, is characterized by an ionization parameter of $1.96<\log\xi <2.05$ and traced mainly by Ne\,{\sc x}, Mg\,{\sc xii}, Si\,{\sc xiii} and Fe\,{\sc xviii}.  

\end{abstract} 

\section{Introduction}\label{sec_in} 
Low-mass X-ray binaries (LMXB) are systems composed of a low-mass ($<1M_{\odot}$) donor star and a compact object, such as a black hole (BH) or a neutron star (NS). Most BH LMXBs, in particular, are transient sources, with outbursts that can last from weeks to months before decaying into quiescence  \citep{tan95,rem06b,cor12,tet16}. During such outbursts, BH LMXBs pass through different accretion states, as indicated by the hysteresis pattern shown in a hardness-intensity diagram \citep{fen04,fen12}. For these systems, it has been observed that winds are stronger in the soft accretion states during which jets are quenched \citep{mil06a,mil06dd,dia07,kub07,ued09bb,pon12,dia14}. The crucial question of why the wind is absent during the hard accretion state remains unanswered. Some possibilities include thermal instabilities \citep{cha13,bia17}, full ionization of the plasma \citep{ued10,dia12,dia14,dia16}, density changes in the wind \citep{mil12a}, geometrical changes in the system \citep{ued10} or mass depletion between winds and jets \citep{nei09}.

The BH~LMXB~IGR~J17091-3624 was discovered by {\it INTEGRAL} in 2003 \citep{kuu03}. During its 2011 outburst the source exhibited a complex variability behavior, traced by the presence of quasi-periodic oscillations (QPOs) with low and high frequencies  \citep{alt11,rod11,alt12}. A notable highly regular flaring heartbeat pattern was identified, which had only been detected previously in the X-ray binary GRS~1915+105 \citep{bel00}. Although similar, \citet{cou17} suggested that the physical processes involved are different for the two sources. Moreover, there was a detection of an ultra fast outflow (UFO) during the outburst \citep{kin12}, which has not been detected since. In February 2016 a new outburst of IGR~J17091-3624 started \citep{mil16c}. \citet{xu17} analyzed {\it NuSTAR} and {\it Swift} observations during the outburst. They concluded that, compared to the previous outburst of 2011, a reflection component is required in order to model the spectra. Their best-fit model indicates an inclination angle of the accretion disc of $\sim 30^\circ -40^\circ$. \citet{wan18} support these conclusions in their analysis of the same data, although their reflection model requires an abundance of Fe=($3.5\pm 0.3$)  compared to  ($1.77$<Fe<$2.95$)   reported by \citet{xu17}.  However, they pointed out that model assumptions, such as the photon-index of the incident continuum spectrum and the density profile, strongly affect the iron abundance.

     \begin{figure*}
   \begin{center}
     \includegraphics[scale=0.38]{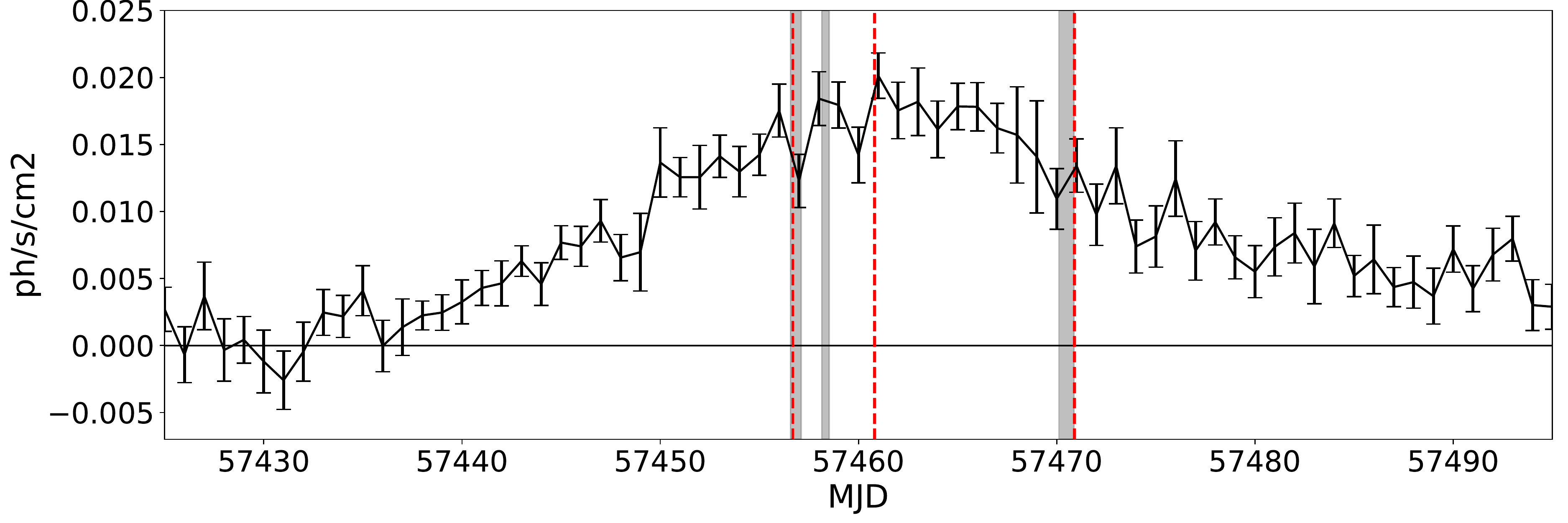}
     \caption{{\it Swift/BAT} daily average lightcurve of the LMXB~IGR~J17091-3624  in the 15--50 keV  energy range. Shaded regions indicate the dates for the {\it XMM-newton} observations while vertical red lines indicate the dates for the {\it ATCA} observations.}\label{fig_swift}
        \end{center}
   \end{figure*}    
   
        \begin{figure*}
   \begin{center}
     \includegraphics[scale=0.55]{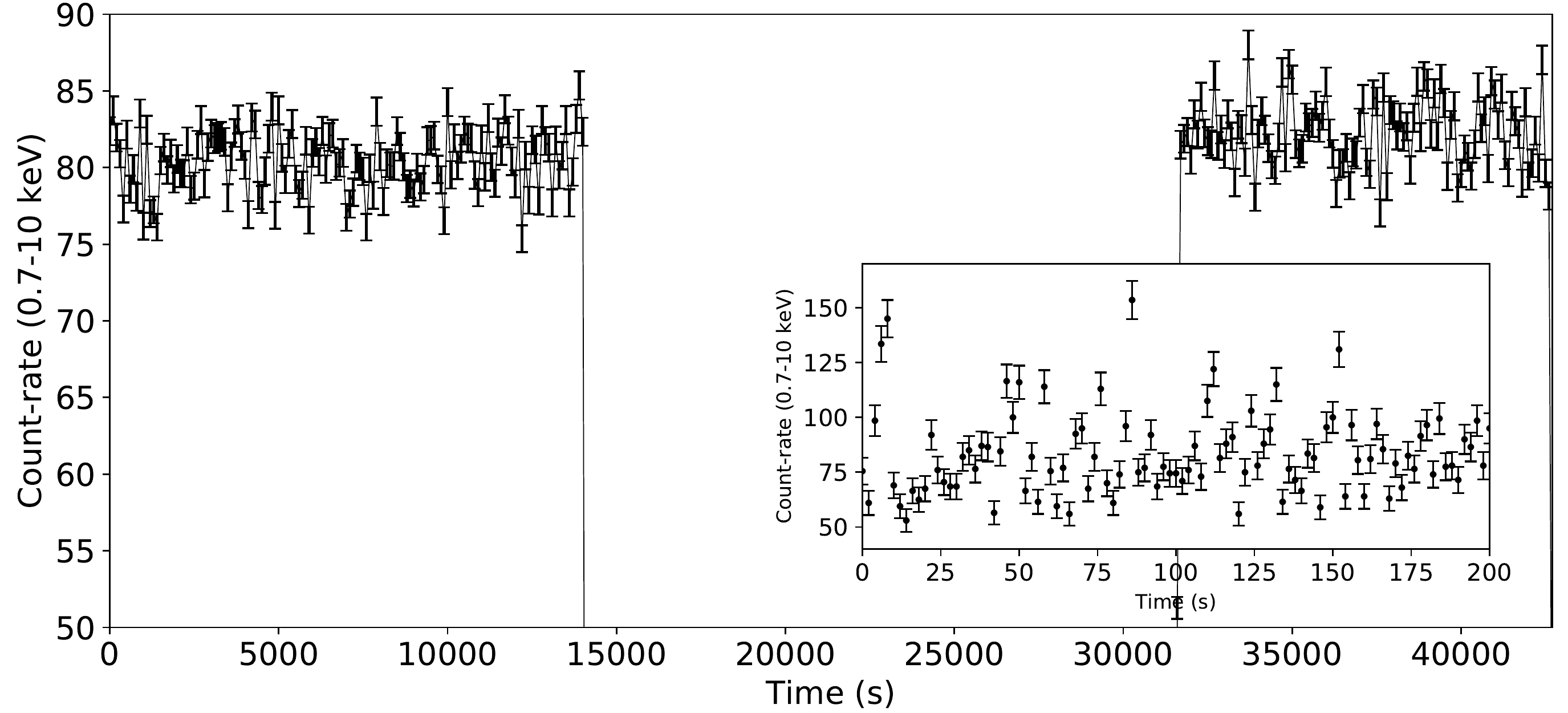}\\
     \includegraphics[scale=0.55]{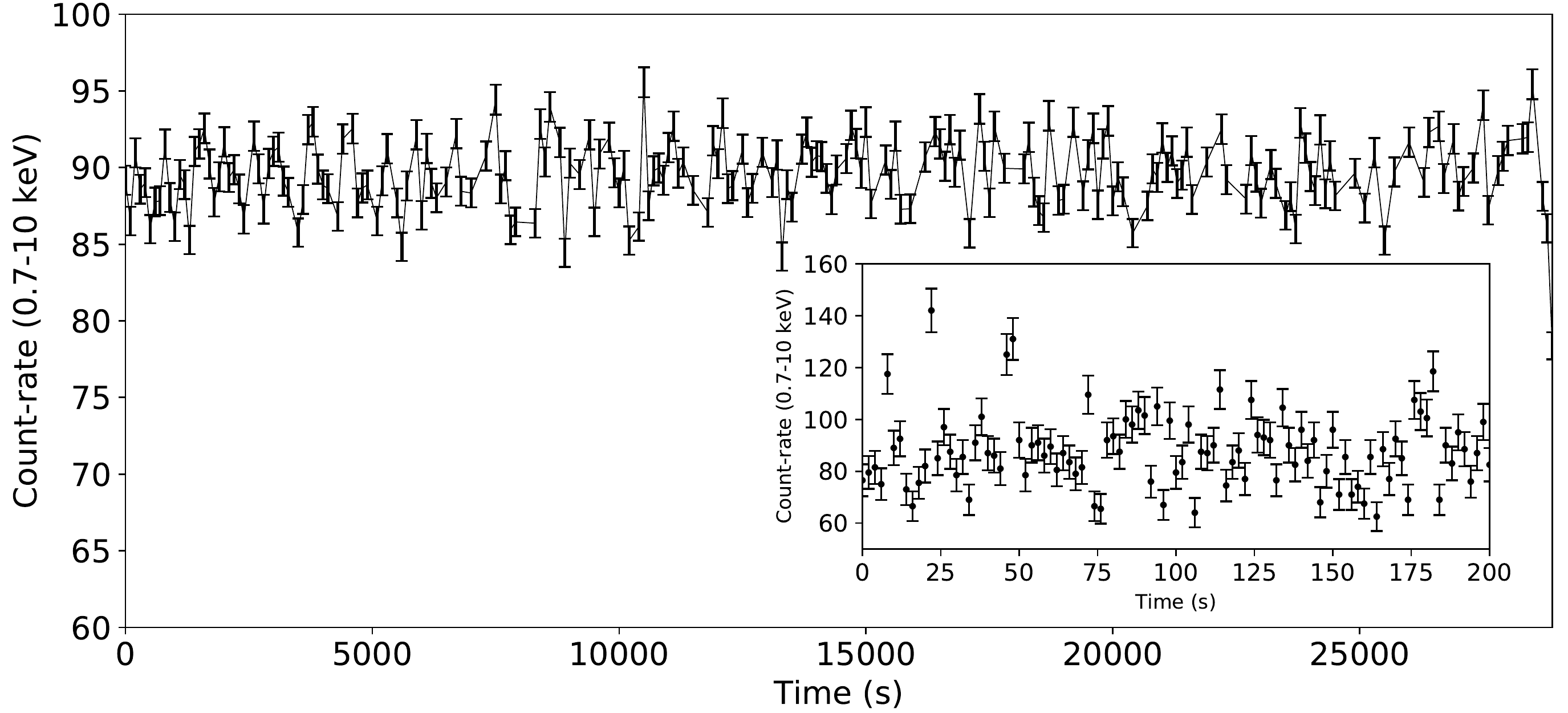}\\
      \includegraphics[scale=0.55]{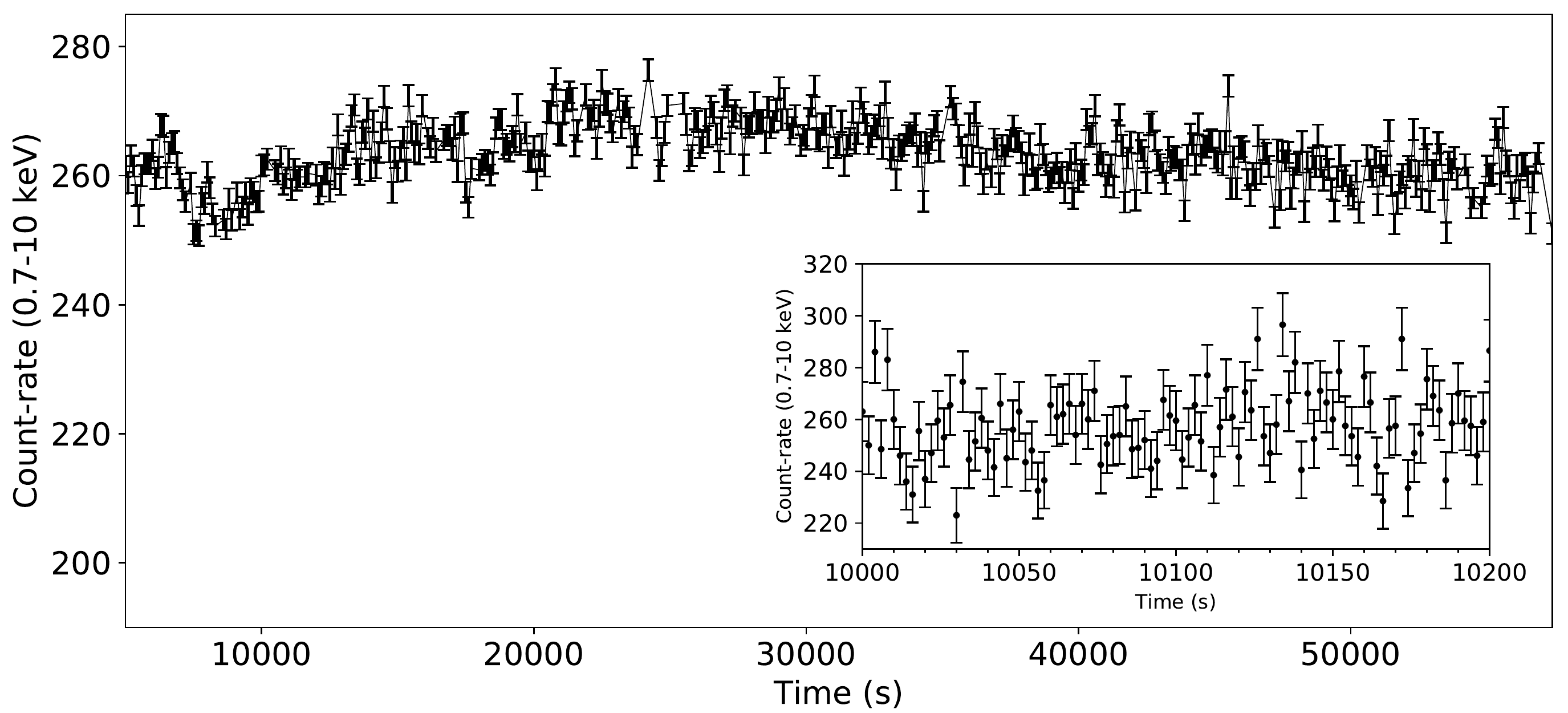}
     \caption{X-ray light curves of the source LMXB~IGR~J17091-3624 as observed by {\it XMM-Newton} EPIC-pn in the 0.7--10 keV for Obs.~1 (top panel), Obs.~2 (middle panel) and Obs.~3 (bottom panel). A zoom of an smaller region is included for each observation.}\label{fig_xmm_lc}
        \end{center}
   \end{figure*}

With the aim of further studying the connection between accretion state, winds and jets across state transitions \citep[e.g. ][]{dia14,gat19a}  and to confirm the existence of a UFO in this source, we triggered simultaneous {\it XMM-Newton} and {\it ATCA} observations of the BH LMXB IGR~J17091-3624. In this paper, we present the results of the analysis of these observations. The outline of this paper is as follows.  In Section~\ref{sec_dat} we describe the observations used and the data reduction process. In Section~\ref{sec_mod} we describe the spectral fitting procedure. In Section~\ref{sec_seds} we analyze the thermal stability curves obtained for all observations. In Section~\ref{sec_dis} we discuss the results obtained. Finally, we summarize the main results of our analysis in Section~\ref{sec_con}. For the spectral analysis we use the {\sc xspec} data fitting package (version 12.9.1p\footnote{\url{https://heasarc.gsfc.nasa.gov/xanadu/xspec/}}). For the X-ray spectral fits, we assumed $\chi^{2}$ statistics in combination with the \citet{chu96} weighting method, which allows the analysis of data in the low counts regime providing a goodness-of-fit criterion \citep[see for example,][]{gat19a}. Errors are quoted at the 90\% confidence level. The abundances are given relative to \citet{gre98}. Finally, we assume a source distance of 17~kpc \citep{rod11}.

         \begin{figure}
   \begin{center}
     \includegraphics[scale=0.42]{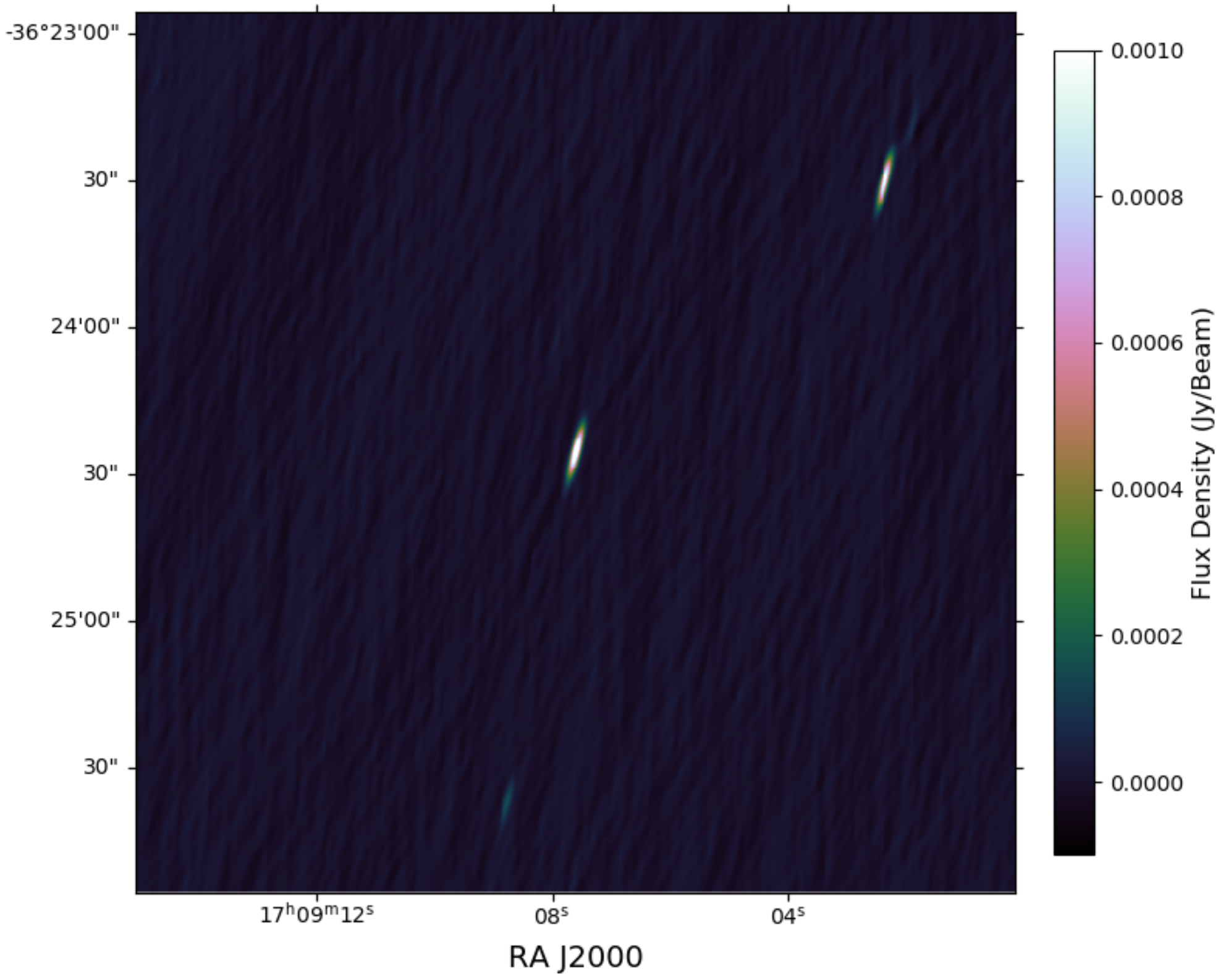}
     \caption{Radio observation obtained with {\it ATCA} in flux density units. The LMXB~IGR J17091-3624 is located in the center.}\label{fig_radio}
        \end{center}
   \end{figure}  
 
 \begin{table}
\footnotesize
\caption{\label{tab_obs}IGR~J17091-3624 observations used in the present work.}
\centering
\begin{tabular}{lccccc}
\hline
Label& Instrument&ObsID & Obs. Date & Exposure     \\
&&&&(ks) \\
\hline
Obs1& {\it XMM-Newton}&0744361501&9 March 2016  & 28   \\   
Obs2& {\it XMM-Newton}&0744361801&11 March 2016  & 31    \\   
Obs3&{\it XMM-Newton}&0744361701&23 March 2016  & 64    \\    
\end{tabular}
\end{table}

 \begin{table*}
\footnotesize
\caption{\label{tab_radio}IGR~J17091-3624 {\it ATCA} observations used in the present work.}
\centering
\begin{tabular}{lcccccc}
\hline
Label& Array & Obs. Date & Exposure &\multicolumn{2}{c}{Flux Density (mJy)} &$\alpha$   \\
&&&(ks)&5.5 GHz&9.0 GHz \\
\hline
Obs.~1& 6B&    9 March 2016   &11 &$1.26\pm 0.01$ &$1.26\pm 0.01$&$0.029\pm 0.027$  \\   
Obs.~2& 6B&    13 March 2016  &5 &$1.50\pm 0.02$&$1.53\pm 0.02$&$0.038\pm 0.033$    \\   
Obs.~3& H214   &23 March 2016  &6   &$1.63\pm 0.05$&$1.90\pm 0.04$&$0.31\pm 0.16$    \\    
\end{tabular}
\end{table*}

\section{Observations and data reduction}\label{sec_dat}
 \subsection{X-ray spectra}
 
 Table~\ref{tab_obs} shows the details of our observations of IGR~J17091-3624.  The {\it XMM-Newton} European Photon Imaging Camera \citep[EPIC,][]{str01} spectra were obtained in timing mode, using the thin filter. Data reduction, including background subtraction, was done with the Science Analysis System (SAS\footnote{https://www.cosmos.esa.int/web/xmm-newton/sas}, version 16.1.0) following the standard procedure to obtain the spectra. We have binned the EPIC data to oversample the instrumental resolution by at least a factor of 3 and to have a minimum of 20 counts per channel. Because the observations were taken in timing read mode, there are no source-free background regions. Given the high source count rates, the real background should not represent more than 1-2\%  \citep{ngc10} and consequently, we chose not to subtract the spectrum extracted from the outer regions of the CCD, which in reality is dominated by source counts. We fitted the EPIC spectra in the 0.7-10 keV energy range, although we ignore the 1.5-3 keV data because of the presence of instrumental features. We note that the background contributes $\approx 2\%$ to the total count rate. In the case of the Reflection Grating Spectrometers \citep[RGS,][]{denh01} we prefer to analyze the spectra without rebinning in order to avoid the loss of data information. The RGS spectra were fitted in the 0.7--2 keV range. We notice that the {\it XMM-Newton} lightcurves do not show dips. 
 
 We note that during the 2016 outburst multiple {\tt NuSTAR} observations of IGR~J17091-3624 were taken, however we note that they were not simultaneous with respect to the {\it XMM-Newton} observations and therefore cannot be used to model the high-energy band. We find in the {\it XMM-Newton} spectra signatures of reflection via the presence of a broad Fe line, similar to those present in the {\it NuSTAR} data \citep{xu17,wan18}. However, in this paper, due to the lack of simultaneous hard X-ray observations, we use a simple Gaussian model to fit the line rather than a reflection component (see Section~\ref{sec_mod}) and focus for the search of absorbers in the spectra.

    \subsection{X-ray light curve}
    Figure~\ref{fig_swift} shows the daily average lightcurve of the LMXB~IGR~J17091-3624  in the 15--50 keV  energy range obtained with the Burst Alert Telescope (BAT) on board the {\it Swift} observatory. Vertical dashed lines indicate the dates for our {\it XMM-Newton} and {\it ATCA} observations (see Table~\ref{tab_obs}). Figure~\ref{fig_xmm_lc} shows a zoom of the {\it XMM-Newton} EPIC-pn lightcurves for all three observations in the 0.7--10 keV energy region. The curves in the large panels have been rebinned to 100 seconds while in the panels with zoom the lighcurves are rebinned to 2 seconds. Note that, for Obs.~1, the observation was not continuously done. Strong variability is clearly identified in all three observations, although they do not show clear heartbeat patterns like those shown in \citet{jan15}.

   \begin{table*}
\scriptsize
\caption{\label{tab_con_gauss}IGR~J17091-3624 EPIC-pn best-fit results obtained including only the ISM X-ray absorption. }
\centering
\begin{tabular}{llcccccc}
\\
Component&Parameter&Obs1&Obs2&Obs3 &Obs1&Obs2&Obs3  \\
\hline
\hline
\\  
&&\multicolumn{3}{c}{Model A1: {\tt tbabs*(powerlaw+diskbb+gauss)}}&\multicolumn{3}{c}{Model A2: {\tt IONeq*(powerlaw+diskbb+gauss)}}\\
{\tt Tbabs} & $N({\rm H})$&$0.90\pm 0.03 $&$ 0.92\pm 0.02   $ &$ 0.95\pm 0.01 $  &$- $&$ -$&$ -$\\
{\tt IONeq} & $N({\rm H})$-neutral &$- $&$ -$&$ -$ &$0.99\pm 0.03 $&$ 0.97\pm 0.03 $ &$1.01\pm 0.01    $   \\
&$N({\rm H})$-warm &$- $&$ -$&$ -$&$ <0.03  $&$<0.03   $ &$ <0.05  $   \\
&$N({\rm H})$-hot &$- $&$ -$&$ -$ &$<0.02   $&$<0.03   $ &$ <0.04  $   \\
{\tt powerlaw}&$\Gamma $ &$ 1.45\pm 0.02 $&$ 1.45\pm 0.01  $ &$ 2.01\pm 0.01 $ &$1.46\pm 0.01   $&$ 1.44\pm 0.01  $ &$ 1.96\pm 0.01  $  \\
&$norm$ &$0.09\pm 0.01 $&$ 0.10\pm 0.01 $ &$0.43\pm 0.01  $ &$ 0.09\pm 0.01  $&$ 0.09\pm 0.01  $ &$ 0.39\pm 0.01 $  \\
{\tt diskbb}&$kT_{in}$ (KeV) &$0.31\pm 0.03  $&$0.31\pm 0.02  $ &$0.70\pm 0.01  $ &$0.29\pm 0.02   $&$0.34\pm 0.04 $ &$ 0.73\pm 0.01  $  \\
&norm$_{dbb}$ &$547_{-330}^{+574}  $&$669_{-340}^{+498}  $ &$ 67\pm 3$ &$ 778_{-404}^{+507}  $&$ 327_{-47}^{+408}  $ &$ 62\pm 3 $  \\
{\tt gaussian}&E (KeV) &$6.95\pm 0.14  $&$6.68\pm 0.08 $ &$6.98\pm 0.07 $ &$ 7.09\pm 0.15  $&$ 6.74\pm 0.10  $ &$6.96\pm 0.08  $  \\
&$\sigma$ (keV) &$0.84\pm 0.20  $&$ 0.52\pm 0.10  $ &$ 0.97\pm 0.04 $ &$0.90\pm 0.20 $&$0.57\pm 0.13   $ &$0.98\pm 0.03 $  \\
&$norm$ &$ (6.4_{-1.8}^{+2.4})\times 10^{-4} $&$ (3.9_{-0.8}^{+1.0})\times 10^{-4}  $ &$ (1.3\pm 0.1)\times 10^{-3} $ &$ (8.2\pm 1.1)\times 10^{-4}  $&$ (3.9\pm 0.9)\times 10^{-4}  $ &$(1.4\pm 0.1)\times 10^{-3}    $  \\ 
Statistic&$\chi^{2}$/d.of.&$   231/125  $&$  195/124 $& $ 639/124 $ &$ 211/123 $   &$ 186/122 $   &$ 407/122 $   \\
&red-$\chi^{2}$&$1.85  $&$ 1.57 $ &$5.15 $ &$ 1.72 $&$  1.52 $ &$ 3.33 $  \\
Flux& erg cm$^{-2}$ s$^{-1}$ &$1.15\pm 0.13  $&$1.28\pm 0.15    $ &$ 5.79\pm 0.52 $ &$1.16 \pm 0.13  $&$1.24 \pm 0.14  $ &$ 5.08\pm 0.45 $  \\
Count-rate&Model  &$0.014\pm 0.003 $&$ 0.016\pm 0.03 $ &$0.009\pm 0.002  $ &$0.014\pm 0.003 $&$ 0.016\pm 0.03 $ &$0.009\pm 0.002  $ \\
(15-50 keV)&{\it Swift/BAT}  &$0.017\pm 0.002   $ &$0.018\pm 0.002 $ &$0.011\pm 0.002 $  &$0.017\pm 0.002   $ &$0.018\pm 0.002 $ &$0.011\pm 0.002 $  \\      
\\
 \hline
  \\     
&& \multicolumn{3}{c}{Model B1: {\tt tbabs*(nthcomp+diskbb+gauss)}}&\multicolumn{3}{c}{Model B2: {\tt IONeq*(nthcomp+diskbb+gauss)}}\\
{\tt Tbabs} & $N({\rm H})$&$0.92\pm 0.03   $&$ 0.94\pm 0.03  $ &$ 0.84\pm 0.01  $&$- $&$ -$&$ -$ \\
{\tt IONeq} & $N({\rm H})$-neutral &$- $&$ -$&$ -$  &$1.01\pm 0.03    $&$ 0.99\pm 0.03   $ &$ 0.91\pm 0.01 $   \\
&$N({\rm H})$-warm &$- $&$ -$&$ -$ &$ <0.03  $&$ <0.04  $ &$ <0.07  $   \\
&$N({\rm H})$-hot  &$- $&$ -$&$ -$ &$ <0.03  $&$ <0.03  $ &$ <0.02 $   \\
{\tt nthcomp}&$\Gamma $ &$1.57\pm 0.02$&$1.57\pm 0.03  $ &$2.03\pm 0.01$ &$1.57\pm 0.01   $&$ 1.57\pm 0.02  $ &$ 2.01\pm 0.01 $  \\
& $kT_e$ (KeV) &$9_{-8}^{+5} $&$8_{-2}^{+6}  $ &$ 1000^{p}  $ &$9_{-8}^{+5} $&$8.4_{-0.1}^{+11.7}   $&$ 1000^{p} $ \\
& $kT_{bb}$ (KeV) &$0.25\pm 0.02 $&$0.25\pm 0.02   $ &$0.57\pm 0.01  $ &$0.24\pm 0.02 $&$ 0.26\pm 0.02 $ &$ 0.59\pm 0.01 $  \\
&norm$_{nhtcomp}$&$0.09\pm 0.01  $&$0.10\pm 0.01   $ &$0.26\pm 0.01   $ &$0.09\pm 0.01 $&$ 0.10\pm 0.01 $ &$0.24\pm 0.01  $  \\
{\tt diskbb} &norm$_{dbb}$&$1476_{-708}^{+1065}  $&$ 1880_{-786}^{+722}   $ &$ 217\pm 6   $ &$ 1783_{-934}^{+1509}  $&$1282_{-630}^{+690}  $ &$ 192\pm 4  $  \\
{\tt gaussian}&E (KeV)&$6.83\pm 0.12  $&$ 6.64\pm 0.04  $ &$6.86\pm 0.08 $ &$6.97\pm 0.08  $&$ 6.70\pm 0.10  $ &$6.90\pm 0.04 $  \\
&$\sigma$ (keV)&$0.80\pm 0.18  $&$ 0.57\pm 0.12   $ &$ 0.95\pm 0.10  $ &$ 0.91\pm 0.17  $&$0.63\pm 0.16 $ &$ 0.95\pm 0.01  $  \\
&$norm$ &$(6.3_{-1.6}^{+2.1})\times 10^{-4}  $&$4.5_{-1.0}^{+1.2})\times 10^{-4} $ &$ (1.1\pm 0.2)\times 10^{-3}   $ &$ (1.2\pm 0.1)\times 10^{-4}  $&$(4.7_{-1.2}^{+1.6})\times 10^{-4}   $ &$(1.3\pm 0.1)\times 10^{-3}    $  \\
Statistic&$\chi^{2}$/d.of. &$ 211/124  $&$  173/123 $& $ 842/124 $ &$ 189/122 $   &$ 157/121 $   &$ 464/122 $   \\
&red-$\chi^{2}$ &$ 1.70$&$ 1.40 $ &$ 6.79$ &$1.55   $&$1.30 $ &$3.80  $  \\
Flux& erg cm$^{-2}$ &$1.11\pm 0.13 $&$1.26\pm 0.15  $ &$ 2.63 \pm 0.21  $ &$1.14\pm 0.13   $&$ 1.21\pm 0.14  $ &$ 2.60\pm 0.21   $  \\
Count-rate&Model  &$0.014\pm 0.003 $&$0.016\pm 0.03   $ &$0.009\pm 0.002  $ &$0.014\pm 0.003 $&$0.016\pm 0.03   $ &$0.009\pm 0.002  $  \\
(15-50 keV)&{\it Swift/BAT}  &$0.017\pm 0.002   $ &$0.018\pm 0.002 $   &$0.011\pm 0.002 $ &$0.017\pm 0.002   $ &$0.018\pm 0.002 $   &$0.011\pm 0.002 $ \\    
\\ 
 \hline
  \\       
\multicolumn{5}{l}{$N({\rm H})$ column densities in units of $10^{22}$cm$^{-2}$.}\\
\multicolumn{5}{l}{Count-rate {\it Swift/BAT} are daily averaged count rates.}\\
\multicolumn{5}{l}{Unabsorbed flux in the 0.0136--13.60 keV range ($\times 10^{-9}$).  }\\ 
\multicolumn{5}{l}{$p$ shows the parameter pegged at the upper limit.}\\  
\end{tabular}
\end{table*}

 \subsection{Radio observations}\label{sec_radio}
We observed IGR~J17091-3624 with the Australia Telescope Compact Array ({\it ATCA}) on 2016 March 9, 13 and 23, under project code C2514.  The array was in its most extended 6~km configuration for the first two epochs, and the compact H214 configuration for the final epoch.  We observed in two 2048-MHz frequency bands, centred at 5.5 and 9.0 GHz.  We used the standard primary calibrator 1934-638 to set the flux scale and calibrate the instrumental frequency response.  To calibrate the complex gains, we used the nearby calibrator 1714-336.  The data were reduced according to standard procedures in the Common Astronomy Software Application \citep[CASA, v5.4.1;][]{mcm07}.  Imaging was carried out using two Taylor terms to model the sky frequency dependence, and using Briggs weighting with a robust factor of 1.  IGR J17091-3624 was significantly detected in all epochs, and was sufficiently distant from other sources in the field that it could be well identified even in the compact configuration on March 23 (see Figure~\ref{fig_radio}).  None of the epochs had sufficient emission in the field for us to apply any self-calibration.  We measured the source flux density by fitting a point source in the image plane. Table~\ref{tab_radio} shows the flux densities obtained, including the spectral indexes ($\alpha$). A jet was detected during all three observations, with a spectral slope indicating the presence of a compact jet increasing its brightness towards Obs.~3.

\section{X-ray spectral modeling }\label{sec_mod}
  Following \citet{gat19a}, we fitted each observation with two phenomenological models to account for different spectral states. The models are (using {\sc xspec} nomenclature):

   \begin{itemize} 
    \item Model A1: {\tt tbabs*(powerlaw+diskbb+gauss)}
    \item Model B1: {\tt tbabs*(nthcomp+diskbb+gauss)}
    \end{itemize}
where {\tt tbabs} is the ISM X-ray absorption model described in \citet{wil00}, {\tt diskbb} corresponds to an accretion disc consisting of multiple blackbody components \citep{mit84,mak86} and {\tt nthcomp} is a model of thermal comptonization that incorporates a low energy rollover, apart from the high energy cutoff \citep{zdz96,zyc99}. The seed photons for the {\tt nthcomp} model are assumed to be from the disk and consequently their temperature is linked to the inner disk radius temperature of the {\tt diskbb} component. In this way, we can account for soft dominated spectra, hard dominated spectra and hybrid cases between them.  For each model, we compare the count rate obtained from the X-ray spectral fits in the 15--50 keV energy range with the {\it Swift}/BAT count rate (i.e. folded with the effective area of the instrument), as a first criterion to consider a valid fit.
\\
\\

          \begin{figure*}
   \begin{center}
     \includegraphics[scale=0.325]{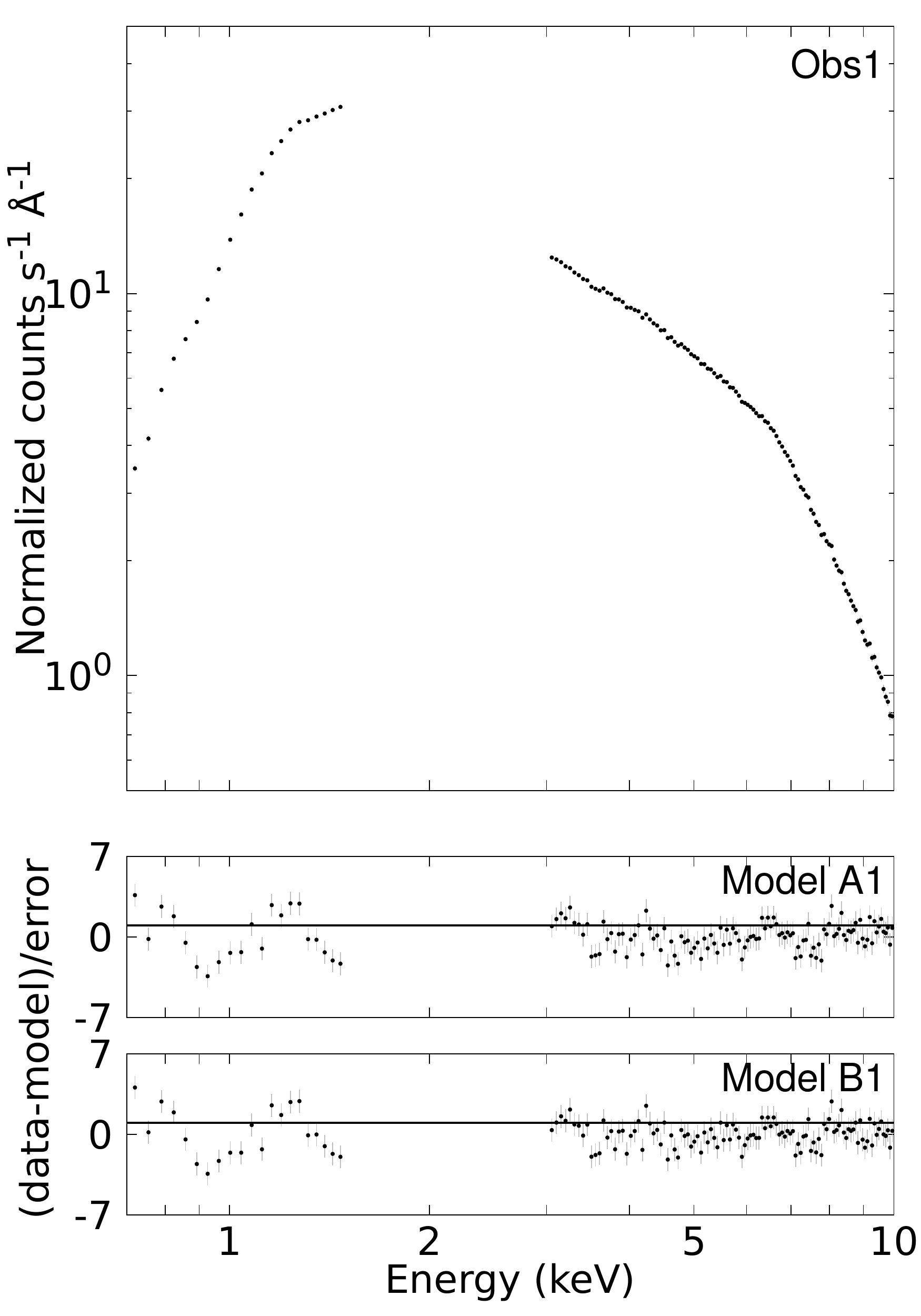} 
         \includegraphics[scale=0.325]{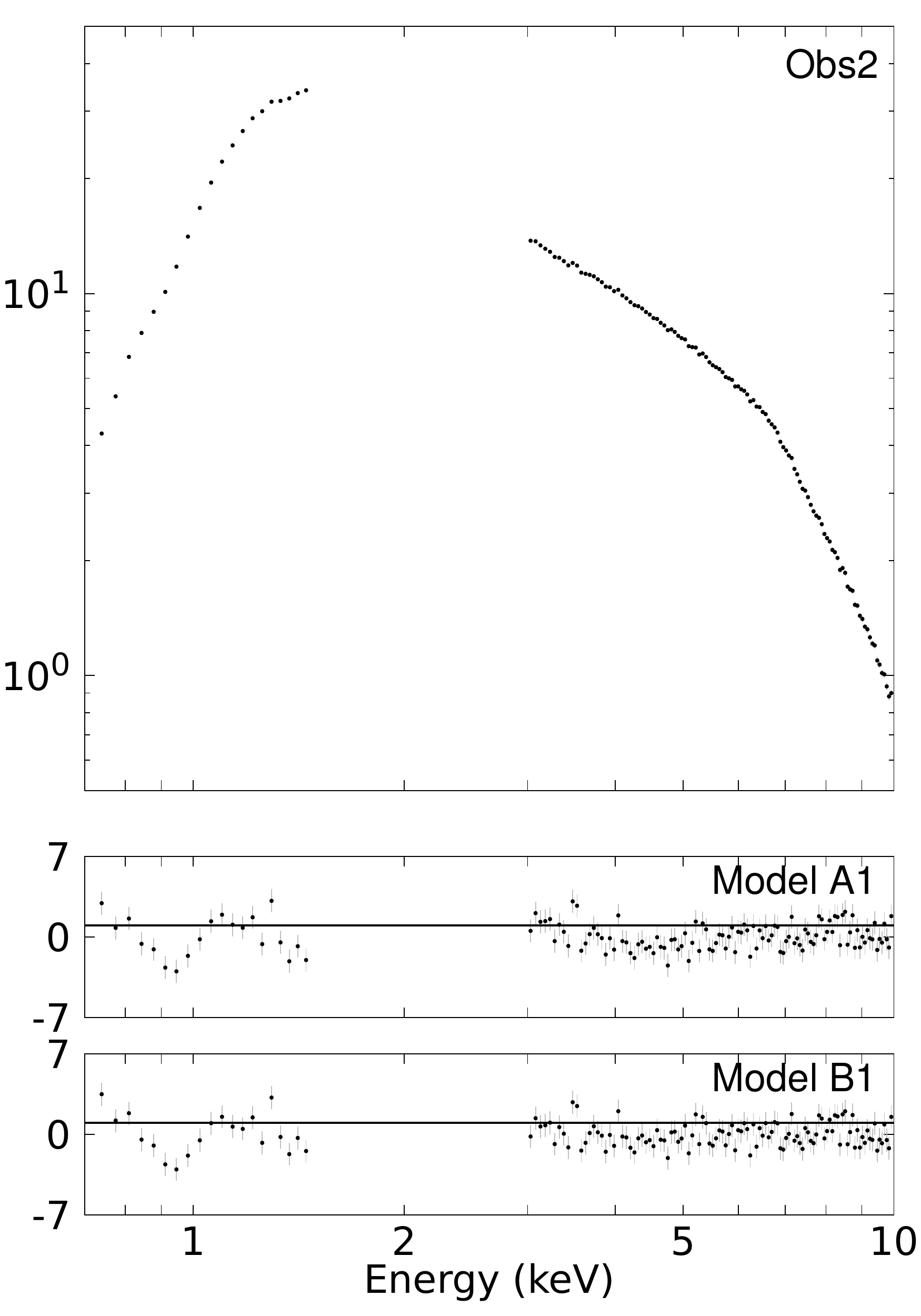} 
          \includegraphics[scale=0.325]{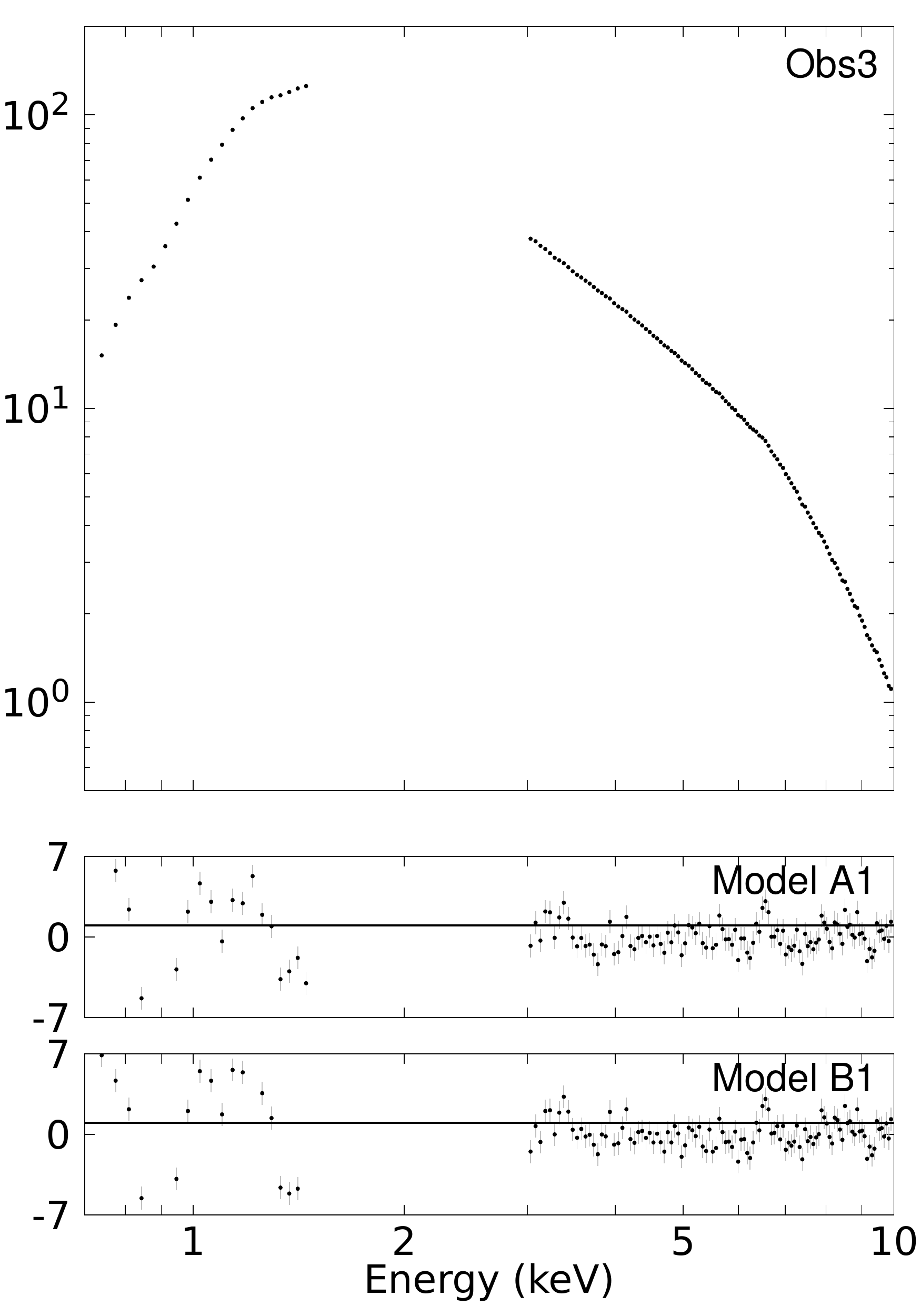} 
     \caption{IGR~J17091-3624 EPIC-pn best-fit obtained for Obs.~1,~2 and 3 using the models described in Table~\ref{tab_con_gauss}. Lower panels show the data/model ratios obtained. }\label{fig_con_obs_1_2_3}
        \end{center}
   \end{figure*}

  \subsection{Continuum fitting}\label{model_first_case}

Figure~\ref{fig_con_obs_1_2_3} shows the best-fit models and residuals obtained from the EPIC-pn spectral fits for all observations while Table~\ref{tab_con_gauss} lists the best-fit parameters.  For both, models A1 and B1, we have included a Gaussian to account for residuals at $\sim 6.8$~keV that remain when fitting the continuum only and that resemble a moderately broad Fe line. For Obs.~1 and 2 there is a general trend that both the disc component temperature and the photon-index are similar between them. Obs.~3, in the other hand, shows a larger photon-index and disc temperature compared to the first observations.  We have estimated the power-law contribution to the total unabsorbed flux in the 2-20 keV energy range to be $>97\%$ for all models used to fit Obs.~1 and 2. For Obs.~3 we have estimated the power-law contribution to be  $\sim 92\%$ (Model~A1) and $\sim 87\%$ (Model~B1), respectively. It has been shown that a power-law contribution $>80\%$ indicates a hard accretion state while a power-law contribution $<25\%$ indicates a soft accretion state \citep{mcc06,bel10}. Using this criterium,  it is clear that Obs.~1 and~2 correspond to a hard accretion state. In the case of Obs.~3, apart from the decreasing of the power-law contribution to the total unabsorbed flux, we note that the both the photon-index and the disc temperature increase. In this sense, Obs.~3 is likely in a hard-intermediate state indicating the start of a spectral state transition \citep[see, for example, ][]{dun10,del16,fur16,shi19}. As Figure~\ref{fig_con_obs_1_2_3} shows, for both models A1 and B1, there are large residuals in the soft part of the spectra and the $\chi^{2}$ values obtained tend to be large, specially for Obs.~3 (see Table~\ref{tab_con_gauss}). Hence, we next try to account for these residuals by including more complex absorption models.

 \begin{table*}
%\scriptsize
\caption{\label{tab_ism}IGR~J17091-3624 EPIC-pn  best-fit results obtained including a {\tt warmabs} component. }
\centering
\begin{tabular}{llccc}
\\
Component&Parameter&Obs.~1&Obs.~2&Obs.~3 \\
\hline
\hline
\\  
\multicolumn{5}{c}{Model A3: {\tt IONeq*warmabs*(powerlaw+diskbb+gauss)}}\\
{\tt IONeq} & $N({\rm H})$-neutral & $ 1.04\pm 0.02 $   &$0.92\pm 0.02 $ &  $ 0.72\pm 0.03 $    \\
&$N({\rm H})$-warm   & $ <0.03   $   &$<0.03   $ &  $0.06\pm 0.03  $    \\
&$N({\rm H})$-hot   & $ <0.03   $   &$ <0.03  $ &  $ 0.011\pm 0.006 $    \\
{\tt powerlaw}&$\Gamma $   & $ 1.47\pm 0.01  $   &$1.43\pm 0.01  $ &  $ 2.01\pm 0.01  $    \\
&$norm$  & $ 0.09\pm 0.01  $   &$ 0.09\pm 0.01 $ &  $0.23\pm 0.01    $    \\
{\tt diskbb}&$kT_{in}$ (KeV)   & $ 0.26\pm 0.01  $   &$ 0.40\pm 0.02  $ &  $ 0.61\pm 0.01 $    \\
&norm$_{dbb}$  & $ 1760_{-752}^{+907}  $   &$ 123\pm 9  $ &  $167\pm 2 $    \\
{\tt gaussian}&E (KeV)  & $ 7.08\pm 0.10   $   &$ 6.74\pm 0.12  $ &  $ 6.97\pm 0.05   $    \\
&$\sigma$ (keV)  & $0.99\pm 0.10  $   &$ 0.60\pm 0.10  $ &  $ 0.96\pm 0.03   $    \\
&$norm$  & $ (8.9\pm 1.1)\times 10^{-4}   $   &$ (3.9\pm 1.1)\times 10^{-4}  $ &  $ (1.18\pm 0.06)\times 10^{-3}  $    \\
{\tt warmabs}& log$($N${\rm H}/10^{22})$   & $ -(1.16_{-0.42}^{+0.16} ) $   &$ -(1.43_{-0.56}^{+0.24})  $ &  $-(0.69\pm 0.06)    $    \\
& log$(\xi)$  & $ 2.18\pm 0.10   $   &$ 2.02\pm 0.11  $ &  $ 2.01\pm 0.01   $    \\
%& $v_{turb}$   &$ 500$ (fixed) &$ 500$ (fixed)  &$ 500$ (fixed)  \\ 
Statistic&$\chi^{2}$/d.of.   & $201/121  $  & $ 174/121  $    & $376/120   $    \\
&red-$\chi^{2}$     & $  1.66   $   &$ 1.43  $ &  $3.13    $    \\  
\\ 
 \hline
  \\     
  \multicolumn{5}{c}{Model B3: {\tt IONeq*warmabs*(nthcomp+diskbb+gauss)}}\\
{\tt IONeq} & $N({\rm H})$-neutral    & $ 0.71\pm 0.10$   &$ 0.73\pm 0.10  $ &  $ 0.69\pm 0.06 $    \\
&$N({\rm H})$-warm  & $<0.03    $   &$ <0.04  $ &  $ 0.07\pm 0.03 $    \\
&$N({\rm H})$-hot  & $  <0.03  $   &$  <0.03 $ &  $ 0.011\pm 0.006 $    \\
{\tt nthcomp}&$\Gamma $  & $1.57\pm 0.02  $   &$1.57\pm 0.01 $ &  $ 2.01\pm 0.01  $    \\
& $kT_e$ (KeV)  & $8.60_{-7.30}^{+4.49}  $   &$ 7.83_{-1.48}^{+4.71} $ &  $1000^{p} $    \\
& $kT_{bb}$ (KeV)   & $0.23\pm 0.02    $   &$ 0.25\pm 0.02  $ &  $ 0.61\pm 0.01  $    \\
& $norm$   & $ 0.09\pm 0.01  $   &$0.10\pm 0.01   $ &  $ 0.23\pm 0.01 $    \\
{\tt diskbb} &norm$_{dbb}$   & $ 1141_{-816}^{+1336}   $   &$880_{-578}^{+643}   $ &  $167\pm 4  $    \\
{\tt gaussian}&E (KeV)  & $ 6.87\pm 0.12   $   &$6.65\pm 0.14   $ &  $6.95\pm 0.01  $    \\
&$\sigma$ (keV)   & $ 0.85\pm 0.16  $   &$ 0.59\pm 0.10  $ &  $ 0.98\pm 0.10   $    \\
&$norm$   & $(6.7_{-1.7}^{+2.5})\times 10^{-4}  $   &$(4.4_{-0.7}^{+0.8})\times 10^{-4}   $ &  $ (1.18\pm 0.03)\times 10^{-3}   $    \\
{\tt warmabs}& log$($N${\rm H}/10^{22})$    & $ -(0.48_{-0.20}^{+0.12})   $   &$-(0.50\pm 0.22)   $ &  $  -(0.79\pm 0.03)  $    \\
& log$(\xi)$   & $ 1.90\pm 0.05   $   &$1.95\pm 0.02   $ &  $ 2.01\pm 0.01   $    \\
%& $v_{turb}$   &$500$ (fixed)     &$500$ (fixed)        &$500$ (fixed)     \\ 
Statistic&$\chi^{2}$/d.of.   & $ 177/120 $   & $ 143/120 $   & $  352/120  $   \\
&red-$\chi^{2}$  & $1.48 $   &$ 1.19  $ &  $ 2.93 $    \\     
 \hline
  \\      
\multicolumn{5}{l}{$N({\rm H})$ column densities in units of $10^{22}$cm$^{-2}$.}\\
\multicolumn{5}{l}{Count-rate {\it Swift/BAT} are daily averaged count rates.}\\
\multicolumn{5}{l}{Unabsorbed flux in the 0.0136--13.60 keV range.  }\\ 
\multicolumn{5}{l}{$p$ shows the parameter pegged at the upper limit.}\\ 
\end{tabular}
\end{table*}

 \begin{table}
%\scriptsize
\caption{\label{tab_warm_3}IGR~J17091-3624  observation 3 RGS best-fit results including the {\tt warmabs} component. }
\centering
\begin{tabular}{llccccccc }
\\ 
Component&Parameter    &   RGS \\ 
\hline 
\hline
 \\ 
   \multicolumn{3}{c}{Model A3: {\tt IONeq*warmabs*(powerlaw+diskbb)}}\\
{\tt IONeq} & $N({\rm H})$-neutral   &$0.91\pm 0.01  $    \\
&$N({\rm H})$-warm   &$<0.04  $    \\
&$N({\rm H})$-hot  &$ <0.03 $      \\
{\tt powerlaw}&$\Gamma$  &$2.01$ (fixed)     \\
&$norm$    &$ 0.04\pm 0.02 $     \\
{\tt diskbb}&$kT_{in}$     &$0.61 $ (fixed)    \\
&norm$_{dbb}$    &$304\pm 9 $    \\
{\tt constant}&   &$ 0.98\pm 0.01 $     \\  
{\tt warmabs}& log$($N${\rm H}/10^{22})$   &$-(1.05\pm 0.10)  $    \\
& log$(\xi)$   &$ 2.02\pm 0.03$    \\
& $v_{turb}$   &$ 60\pm 15$   \\ 
Statistic&$\chi^{2}$/d.of.    &$2732/1864  $    \\
&red-$\chi^{2}$    &$1.47  $     \\   
\\ 
 \hline
  \\    
   \multicolumn{3}{c}{Model B3: {\tt IONeq*warmabs*(nthcomp+diskbb)}}\\
{\tt IONeq} & $N({\rm H})$-neutral    &$ 0.87\pm 0.02 $     \\
&$N({\rm H})$-warm  &$<0.03  $    \\
&$N({\rm H})$-hot    &$<0.03  $     \\
{\tt nthcomp}&$\Gamma$    &$2.01   $ (fixed)  \\
& $kT_e$     &$1000   $ (fixed)   \\
& $kT_{bb}$    &$0.61 $ (fixed)    \\
& $norm$     &$0.40\pm 0.01 $    \\
{\tt diskbb}&norm$_{dbb}$    &$ <0.01 $    \\
{\tt constant}&       &$0.98\pm 0.01 $     \\  
{\tt warmabs}& log$($N${\rm H}/10^{22})$      &$-(0.92\pm 0.01)  $      \\
& log$(\xi)$     &$1.98\pm 0.02  $   \\
& $v_{turb}$      &$ 50\pm 15$    \\ 
Statistic&$\chi^{2}$/d.of.    &$2948/1864  $      \\
&red-$\chi^{2}$     &$1.58  $      \\  
\hline
 \\ 
\multicolumn{3}{l}{{\tt constant} is for RGS order 2. }\\
\multicolumn{3}{l}{$N({\rm H})$ column densities in units of $10^{22}$cm$^{-2}$.}\\
\multicolumn{3}{l}{Count-rate {\it Swift/BAT} are daily averaged count rates.}\\ 
\end{tabular}
\end{table}

                \begin{figure*}
   \begin{center}
             \includegraphics[scale=0.43]{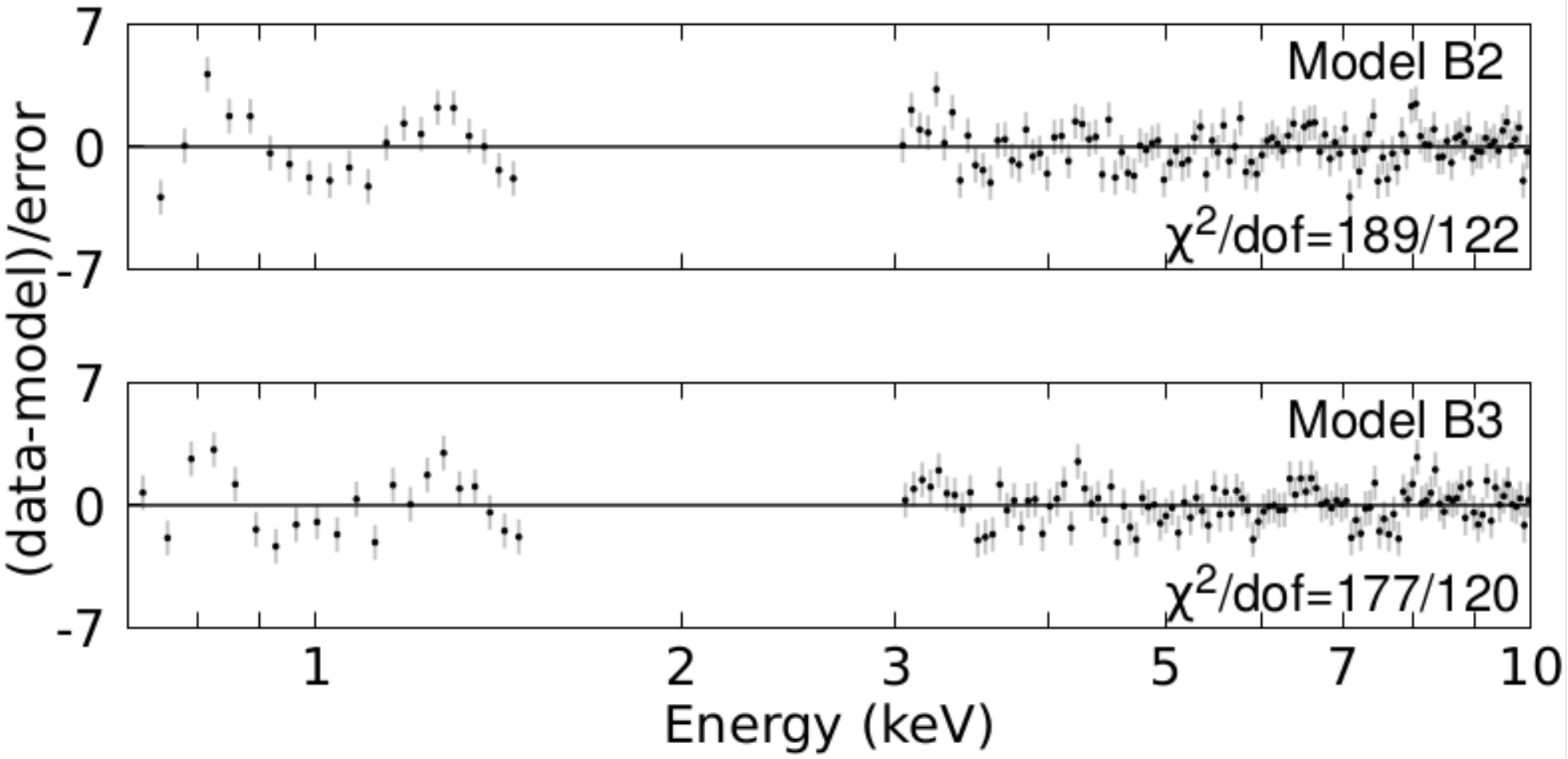}  
             \includegraphics[scale=0.43]{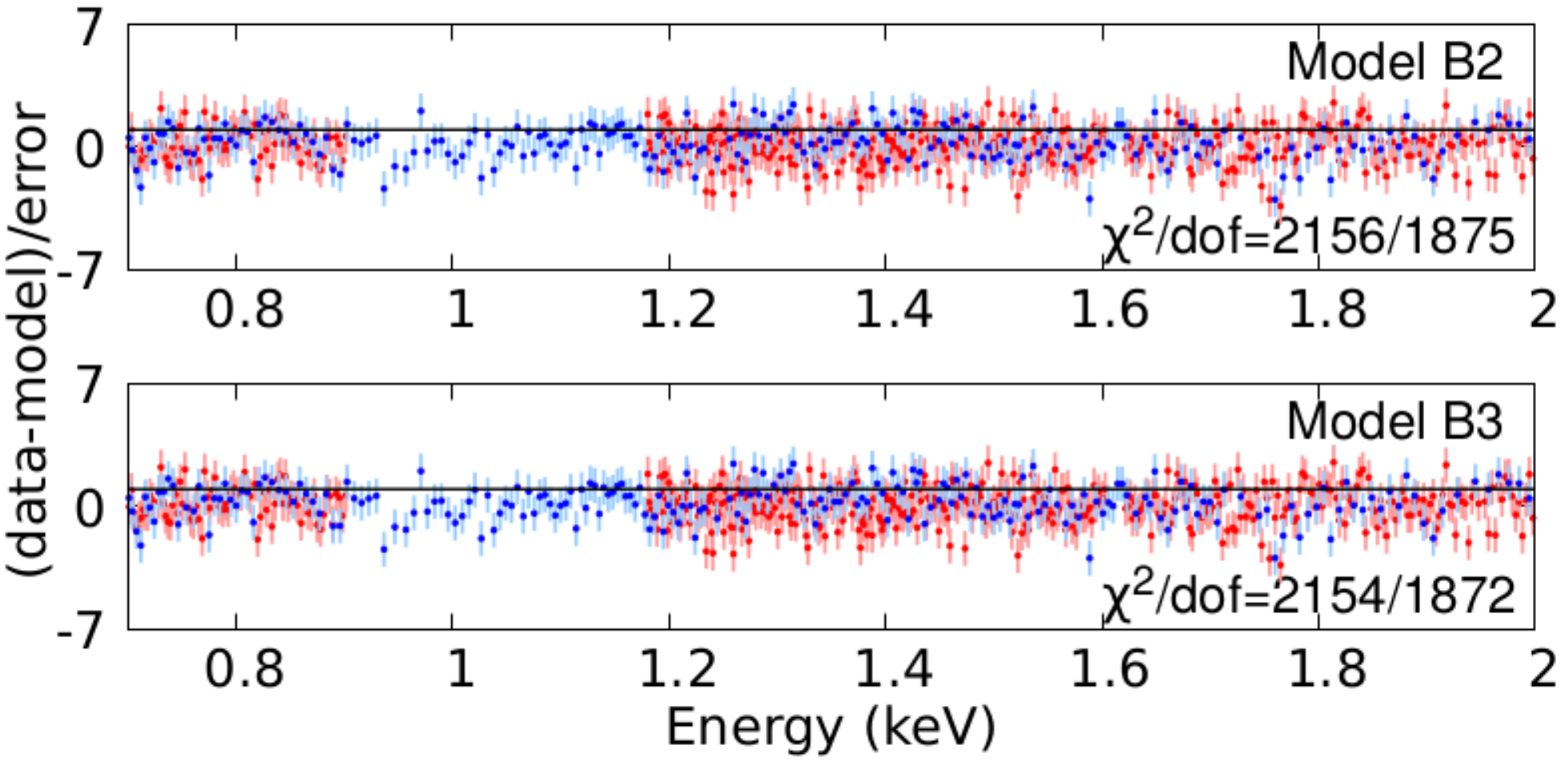} \\
         \includegraphics[scale=0.43]{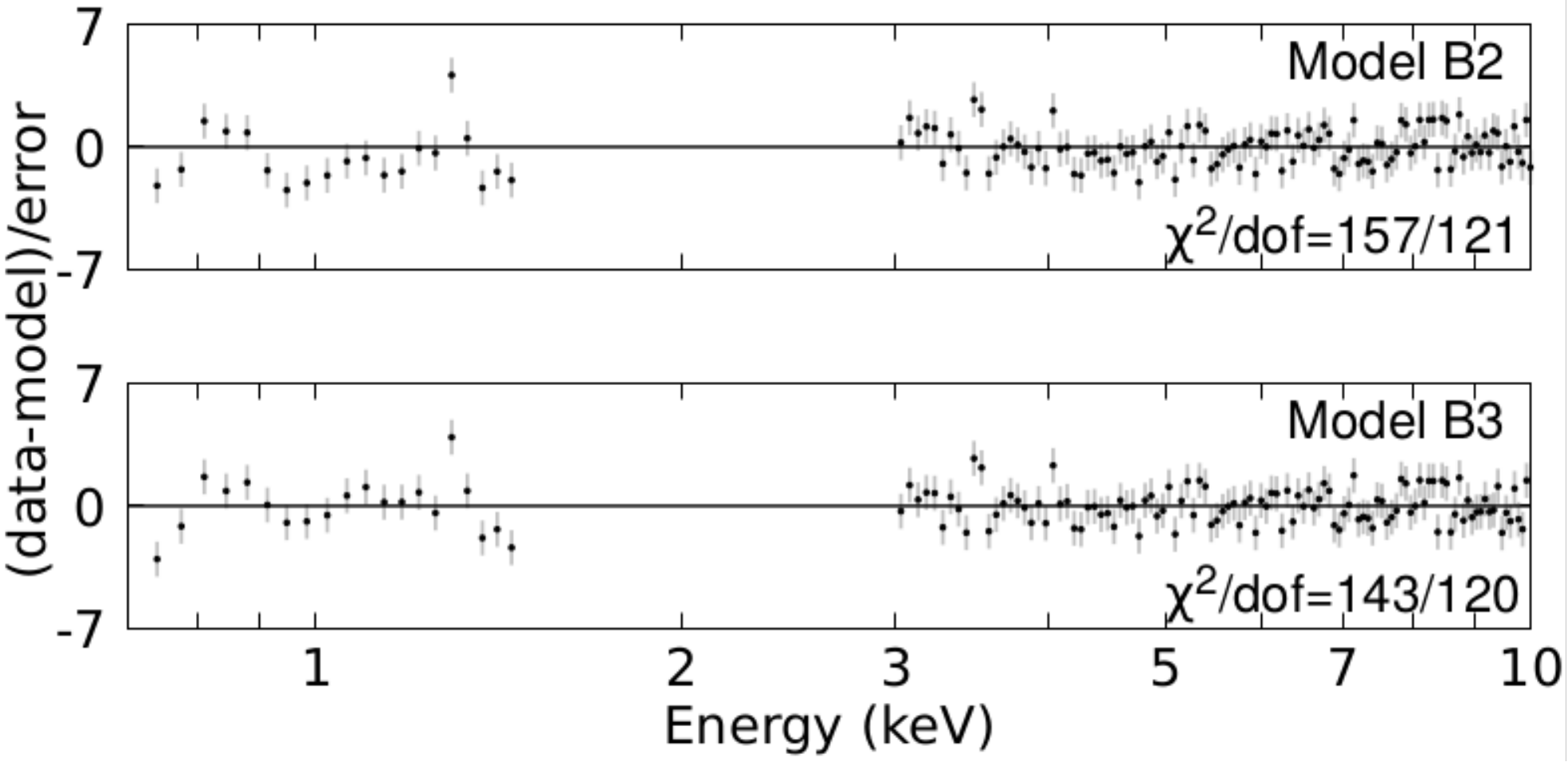} 
         \includegraphics[scale=0.43]{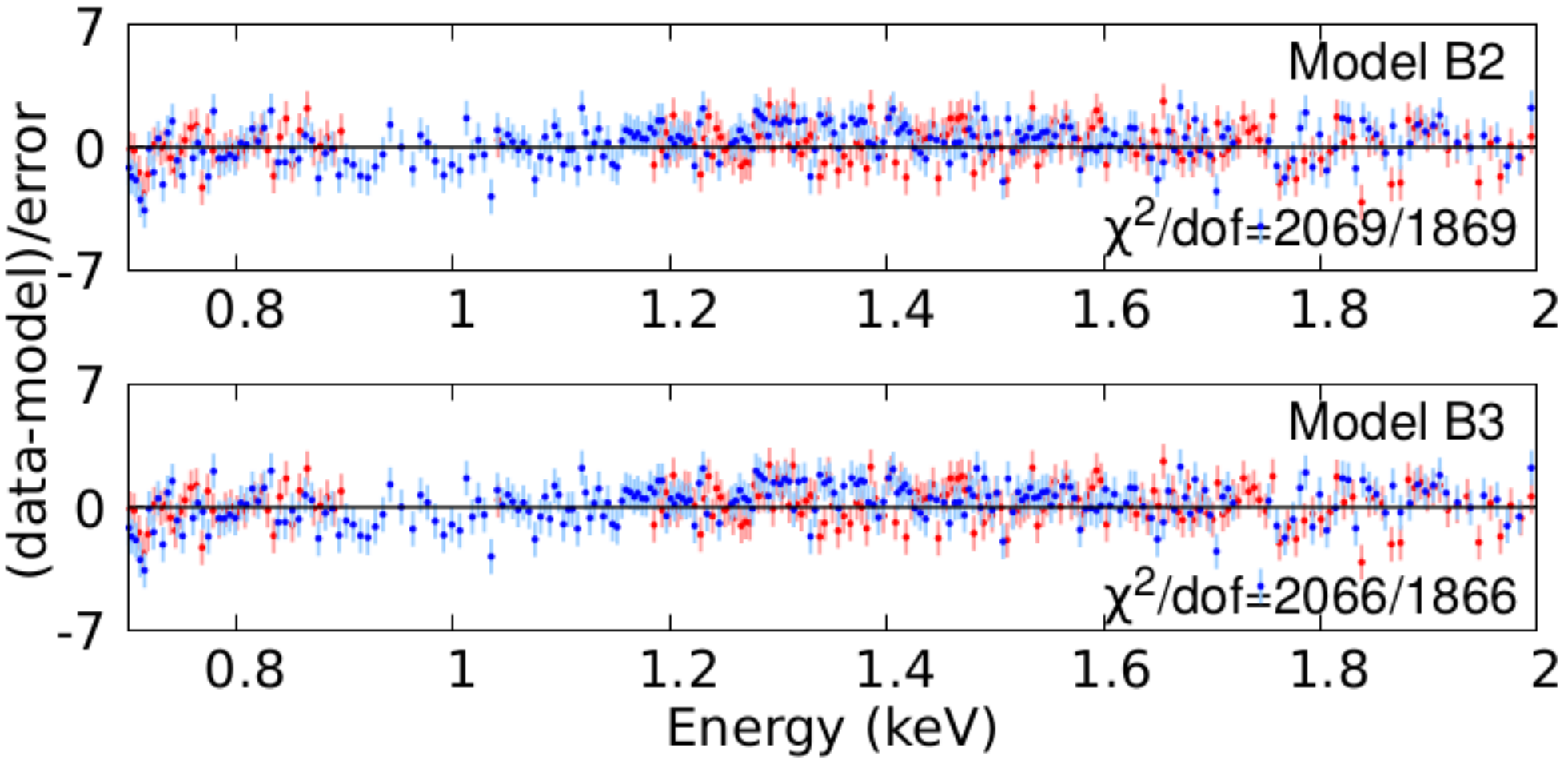} \\
          \includegraphics[scale=0.43]{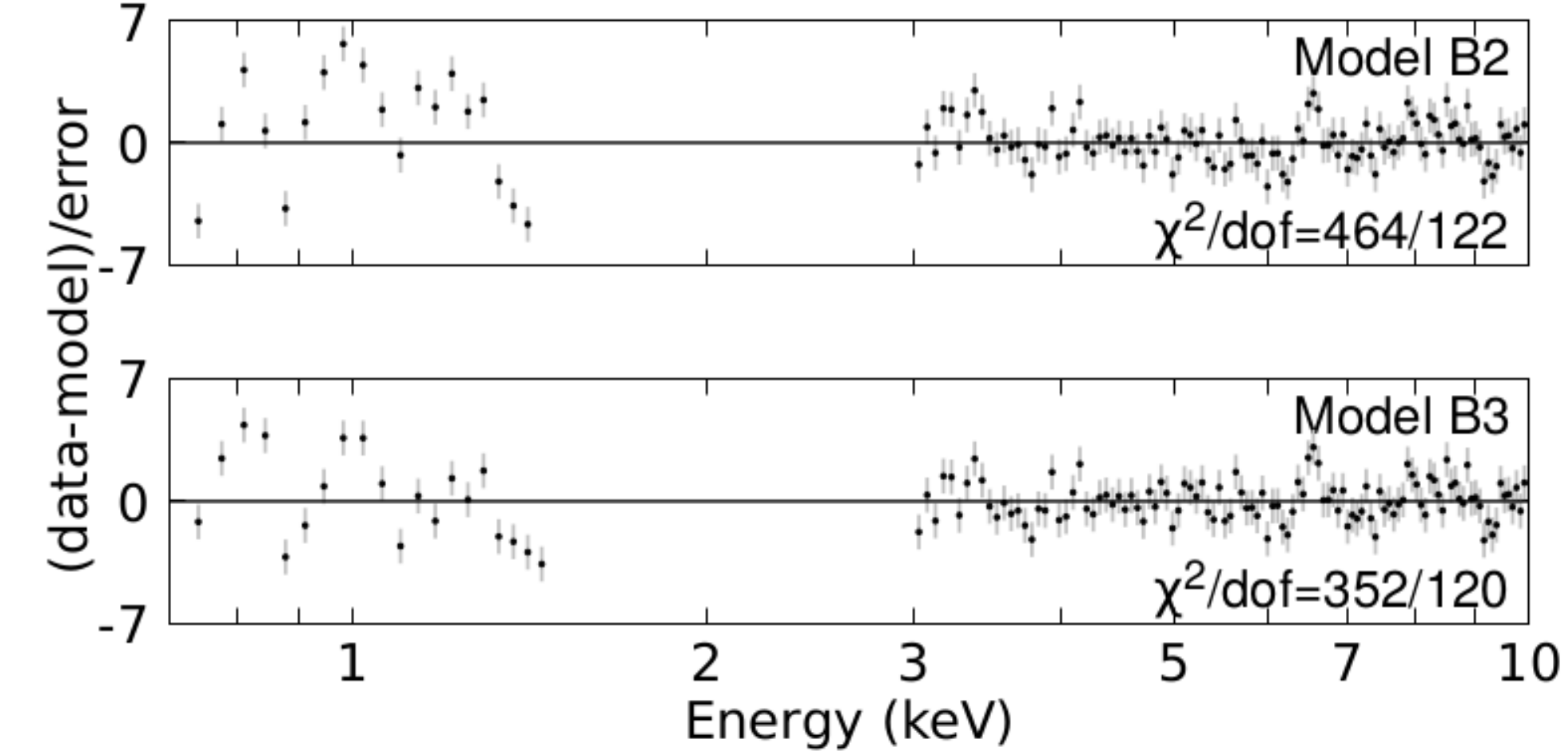} 
         \includegraphics[scale=0.43]{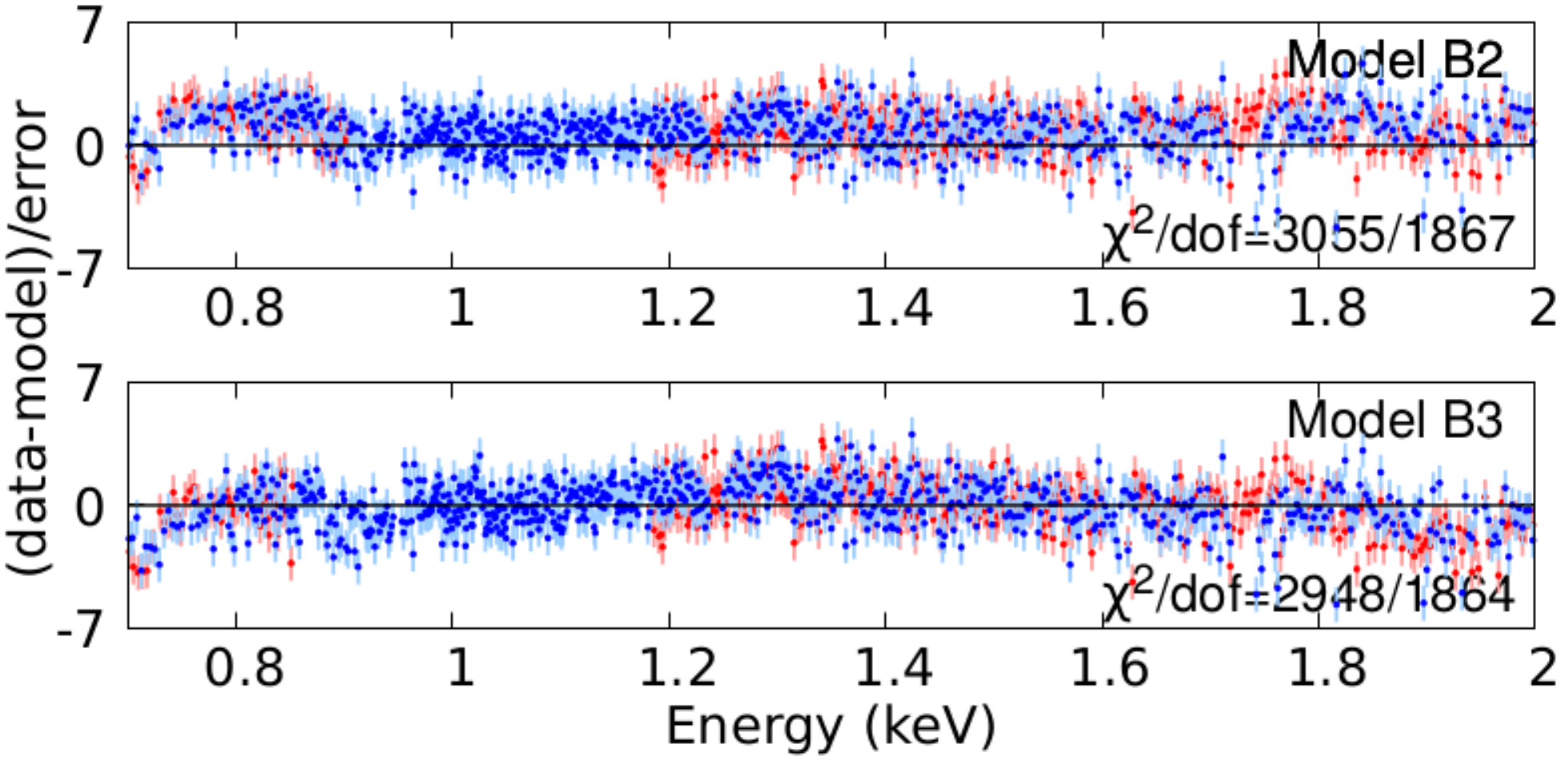} \\
     \caption{IGR~J17091-3624 Epic-pn and RGS best-fit residual using model~B2 and model~B3 (i.e., with and without the {\tt warmabs} component) for Obs.~1 (top panels), Obs.~2 (middle panels) and Obs.~3 (bottom panels). Left panels show the EPIC-pn spectra while right panels show the RGS spectra order 1 (red) and 2 (blue). The statistic is also indicated in units of $\chi^{2}/d.o.f$.  }\label{fig_warm_1_2_modeld}
        \end{center}
   \end{figure*}  
 
                  \begin{figure}
   \begin{center}
             \includegraphics[scale=0.43]{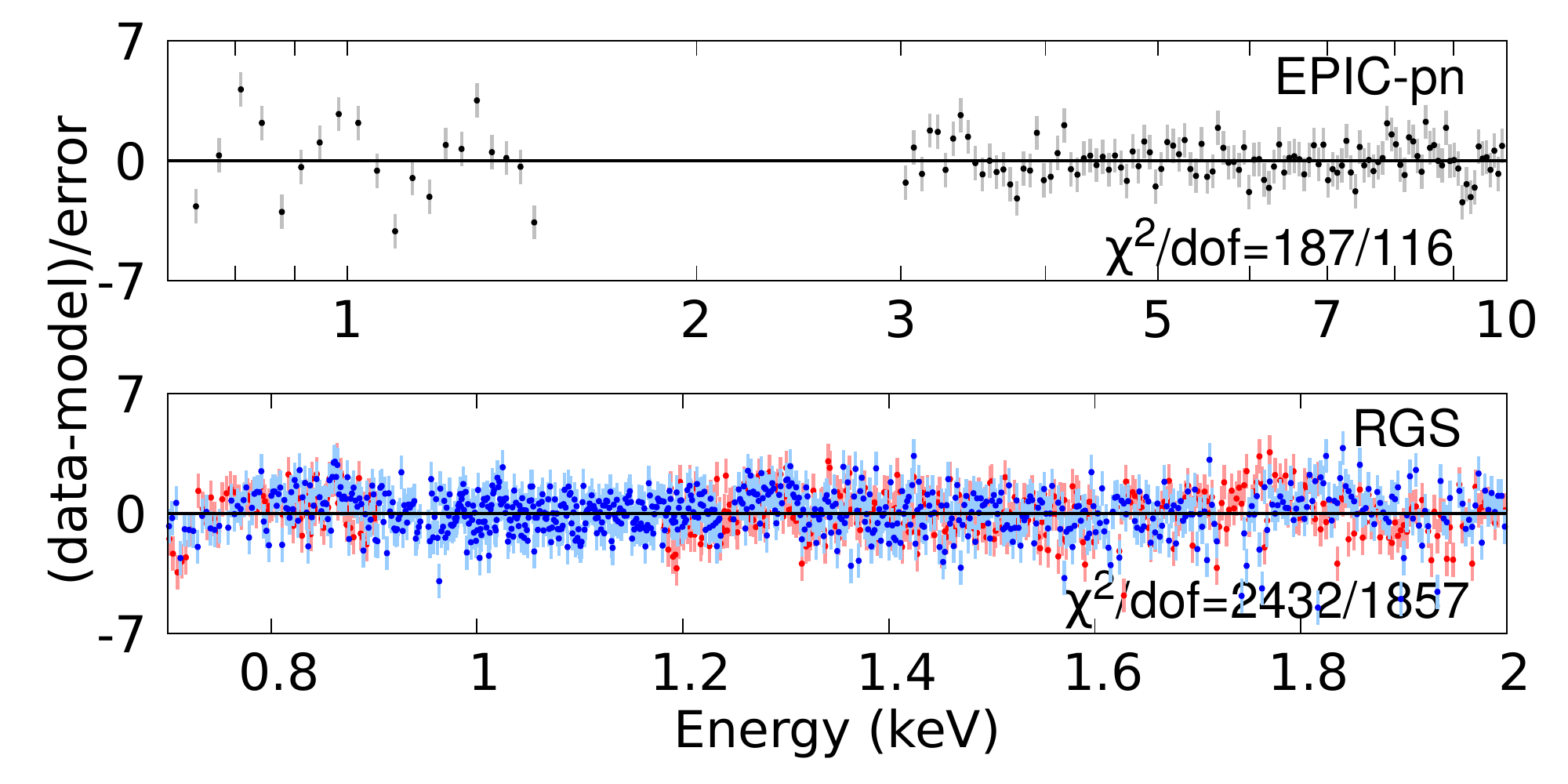}  
     \caption{IGR~J17091-3624 Epic-pn and RGS best-fit residual using model~B3 but allowing variation in temperatures for the {\tt IONeq} component and in the continuum for the RGS data. The statistic is also indicated in units of $\chi^{2}/d.o.f$.}\label{fig_best_fit}
        \end{center}
   \end{figure}

 \begin{figure*}
   \begin{center}    
            \includegraphics[scale=0.43]{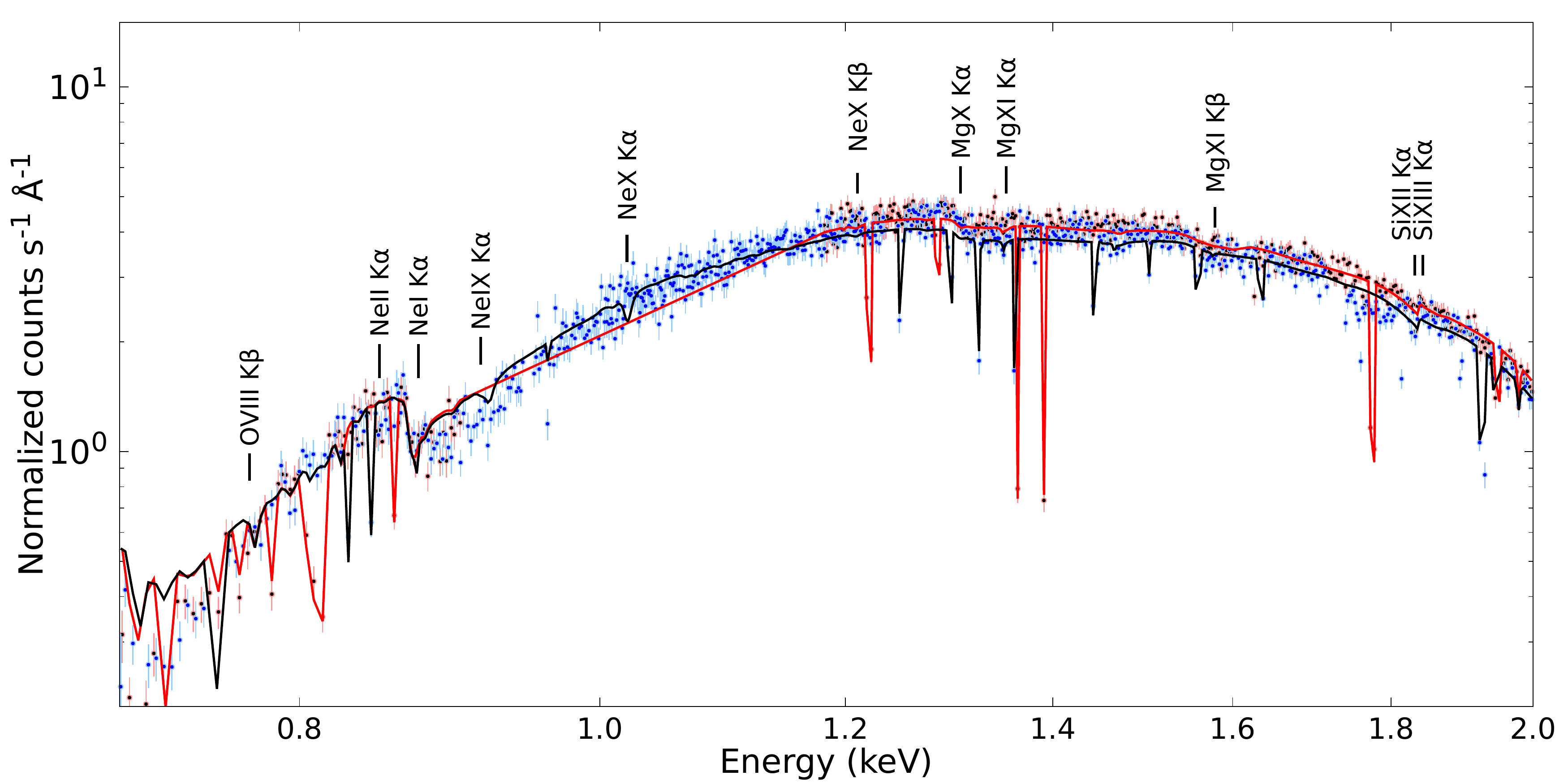}\\
             \includegraphics[scale=0.43]{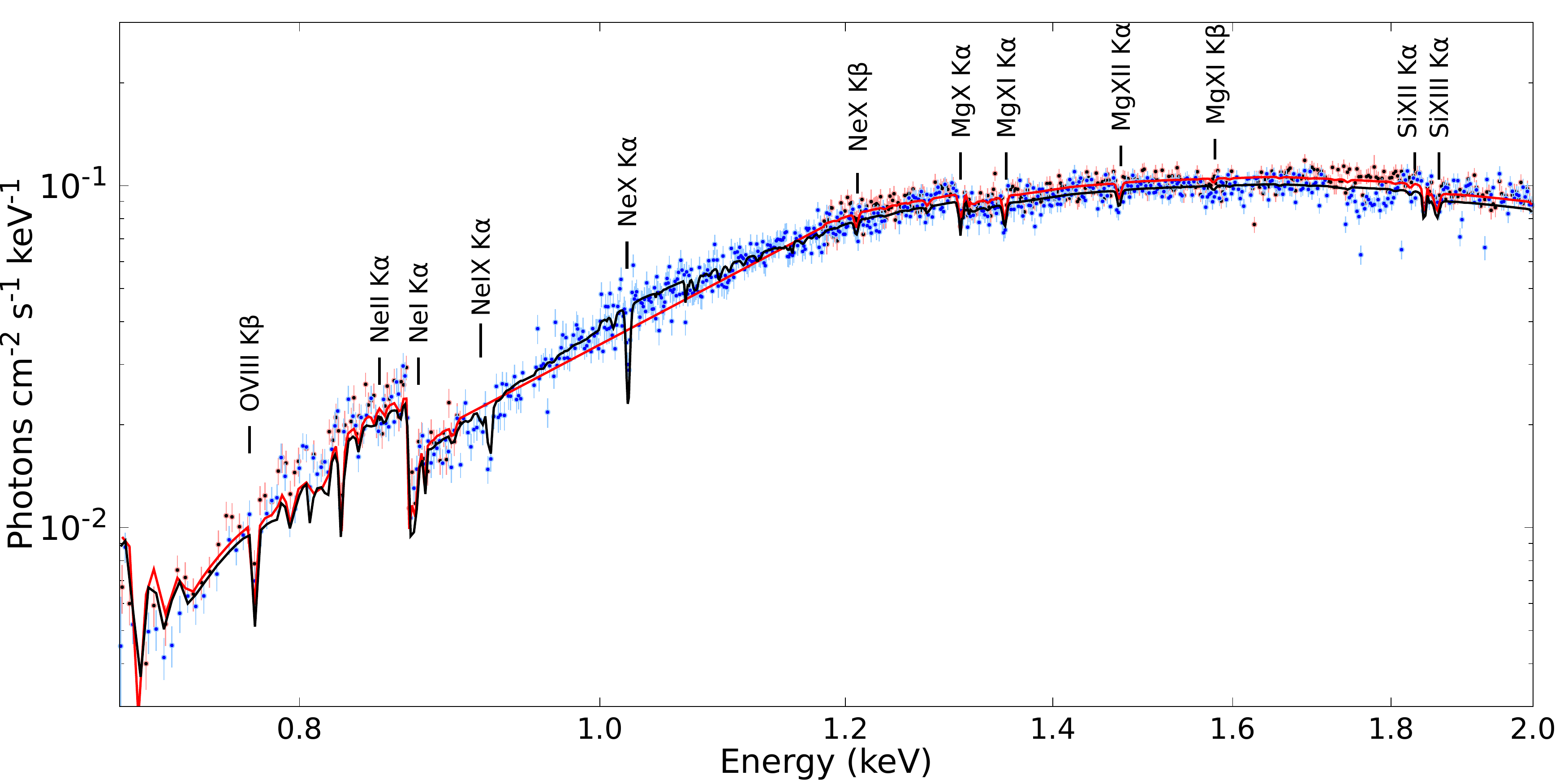} 
     \caption{IGR~J17091-3624 RGS order 1 best-fit results for Obs.~3 using Model~B3 in counts (top panel) and flux (bottom panel) units for RGS 1 (red points) and RGS 2 (blue points). The main absorption lines identified in the spectra are indicated.}\label{fig_warm_3_flux}
        \end{center}
   \end{figure*}

  \begin{table} 
\caption{\label{tab_warmabs_ions}{\tt warmabs} column densities obtained with models A3 and B3 for Obs.~3 (See Section~\ref{warm_obs3} and Figure~\ref{fig_warm_3_flux}). }
%\scriptsize 
\centering   
\begin{tabular}{lcccc}  
\hline
Source &   Model A3&   Model B3\\ 
 \hline
\hline
\\ 
{\it N}(Ne\,{\sc ix})& $ 10.9  \pm 2.1     $ &$ 14.1  \pm  3.3 $    \\ 
{\it N}(Ne\,{\sc x}) & $ 1.8  \pm 0.4  $ &$ 1.7  \pm  0.4   $    \\ 
{\it N}(Mg\,{\sc x}) &$ 0.9  \pm 0.2   $ &$ 1.0  \pm  0.2  $    \\ 
{\it N}(Mg\,{\sc xi}) & $ 8.4  \pm 2.1   $ &$7.9   \pm   1.3 $    \\ 
{\it N}(Mg\,{\sc xii}) & $16.2   \pm 3.1   $ &$ 22.9  \pm   4.5 $    \\ 
{\it N}(Si\,{\sc xii}) & $2.9  \pm 0.6   $ &$ 3.4 \pm   0.5 $    \\ 
{\it N}(Si\,{\sc xiii})& $ 13.1  \pm 2.7   $ &$ 14.4  \pm  2.6  $   \\ 
{\it N}(Si\,{\sc xiv}) & $ 9.4  \pm 1.7   $ &$ 15.9  \pm   3.4 $    \\ 
{\it N}(S\,{\sc xiv}) & $ 1.7  \pm  0.4  $ &$  2.3 \pm    0.4$   \\  
{\it N}(S\,{\sc xv}) & $  2.5 \pm 0.2   $ &$  3.9 \pm    0.6$    \\ 
{\it N}(S\,{\sc xvi}) & $ 0.8  \pm  0.2  $ &$  1.9 \pm    0.4$    \\ 
{\it N}(Fe\,{\sc xviii}) & $ 7.9 \pm  1.3  $ &$ 9.8  \pm    1.9$    \\ 
{\it N}(Fe\,{\sc xix}) & $ 7.2 \pm  1.3  $ &$ 11.2  \pm    1.8$    \\ 
{\it N}(Fe\,{\sc xx}) & $1.9  \pm  0.4  $ &$  3.7 \pm    0.9$   \\  
\hline
\multicolumn{3}{l}{ Column densities in units of $10^{15}$cm$^{-2}$.}   
 \end{tabular}
\end{table}

  \begin{table} 
\caption{\label{tab_gauss_lines}Equivalent width comparison for the main highly ionized absorption lines for Obs.~3.}
%\scriptsize 
\centering   
\begin{tabular}{lccc}  
\hline
Line & Parameter&Theoretical$^{a}$ &  Value\\ 
 \hline
\hline
\\  
 O\,{\sc viii} K$\beta$&E(keV) & 0.78  &$ 0.79\pm 0.01 $ \\ 
&EW (eV) & - &$0.12\pm 0.03$    \\  
 Ne\,{\sc ix} K$\alpha$&E(keV) & 0.92  & $0.92\pm 0.01$  \\ 
&EW (eV) & - & $5.31\pm 1.06$  \\  
 Ne\,{\sc x} K$\alpha$&E(keV) & 1.02 & $1.01\pm 0.01$  \\ 
&EW (eV) & -  &$0.45\pm 0.09$   \\
 Mg\,{\sc x} K$\alpha$&E(keV) & 1.34  & $1.33\pm 0.01$   \\ 
&EW (eV) & - & $0.17\pm 0.03$  \\
 Mg\,{\sc xi} K$\alpha$&E(keV) & 1.35  & $1.34\pm 0.01$   \\ 
&EW (eV) & -  &$0.04\pm 0.02$  \\
 Mg\,{\sc xii} K$\alpha$&E(keV) & 1.47 &$1.47\pm 0.01$      \\ 
&EW (eV) &  -  & $0.44\pm 0.13$  \\
 Si\,{\sc xii} K$\alpha$&E(keV) & 1.85 & $1.84\pm 0.01$     \\ 
&EW (eV) & -  & $<0.03$   \\
 Si\,{\sc xiii} K$\alpha$&E(keV) & 1.87 &$1.86\pm 0.01$      \\ 
&EW (eV) & -  & $<0.04$   \\
\hline    
\multicolumn{4}{l}{$^{a}$Theoretical values taken from {\sc xstar}.}   
 \end{tabular}
\end{table}

         \begin{figure}
   \begin{center}
     \includegraphics[scale=0.43]{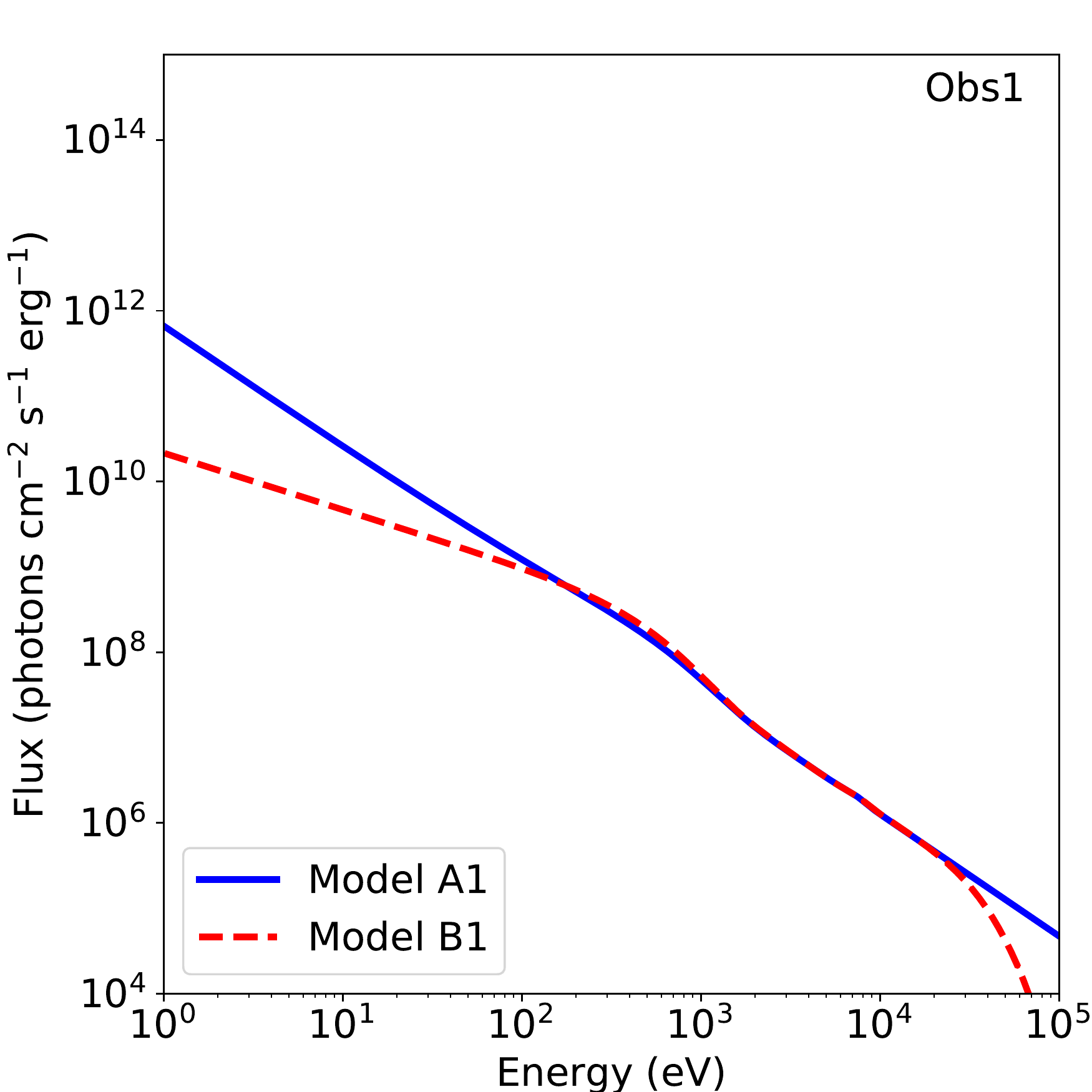} \\
         \includegraphics[scale=0.43]{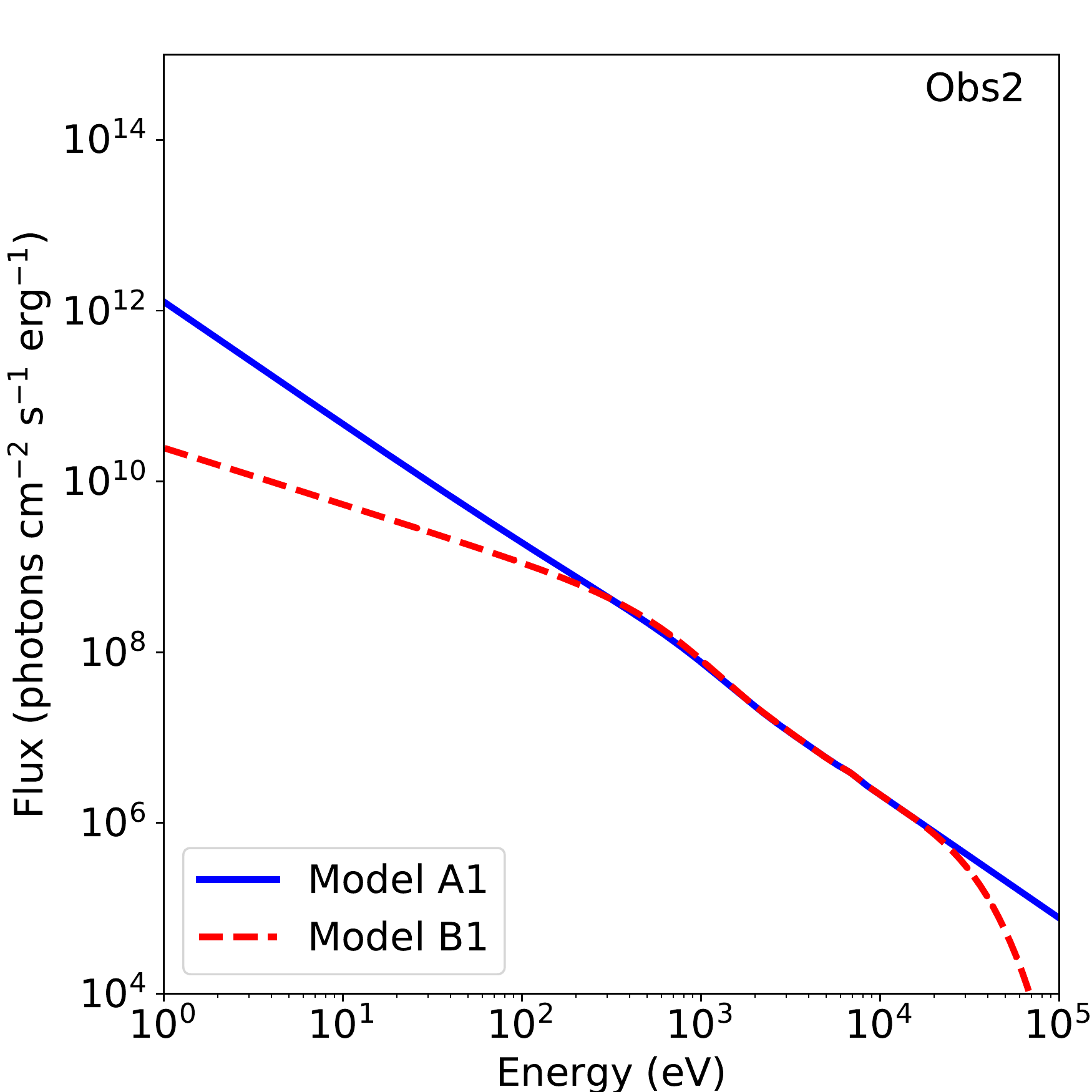} \\
             \includegraphics[scale=0.43]{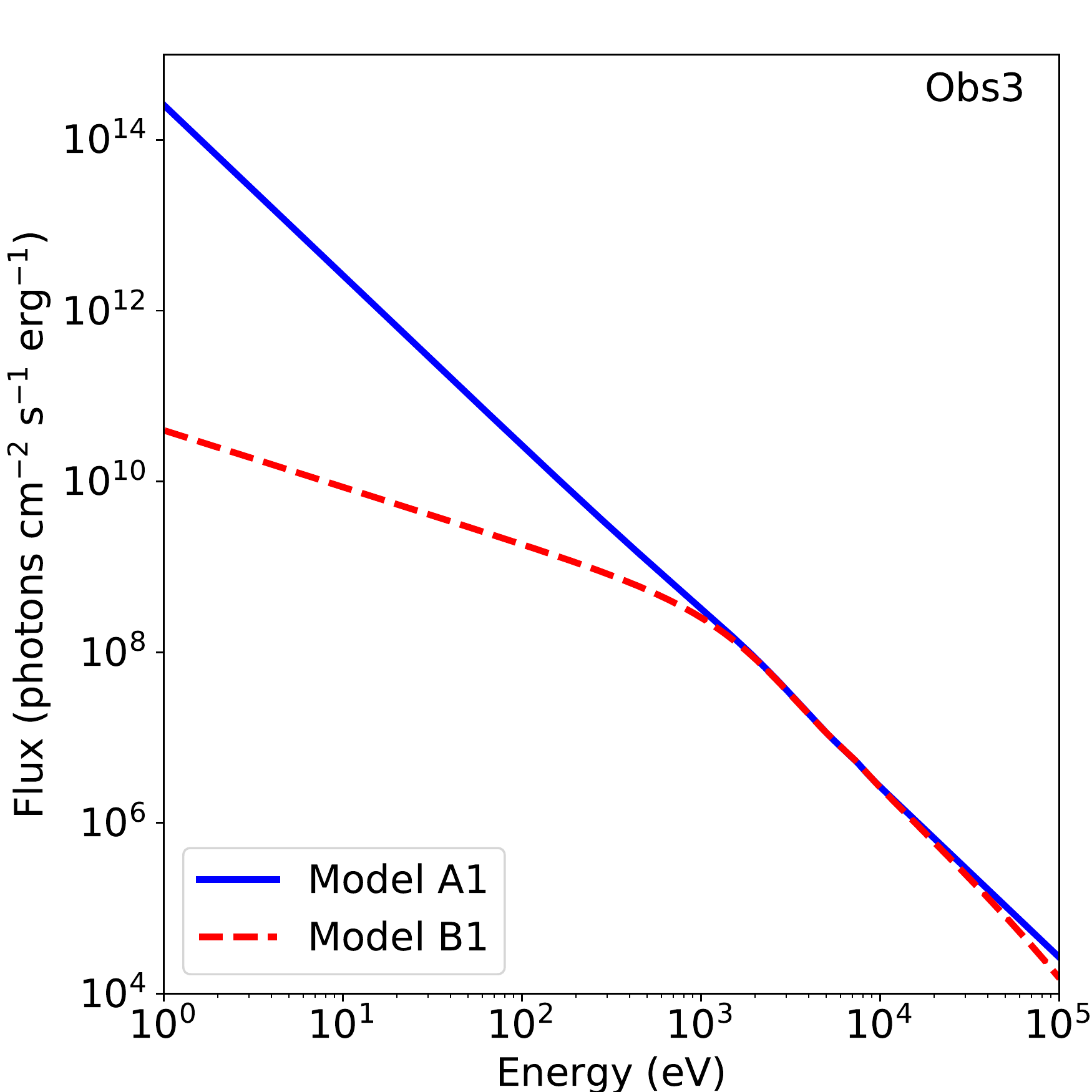}   
     \caption{Spectral energy distributions obtained from the continuum modeling for all observations (see Section~\ref{sec_seds}).}\label{fig_seds}
        \end{center}
   \end{figure}  
   
                \begin{figure}
   \begin{center}
     \includegraphics[scale=0.42]{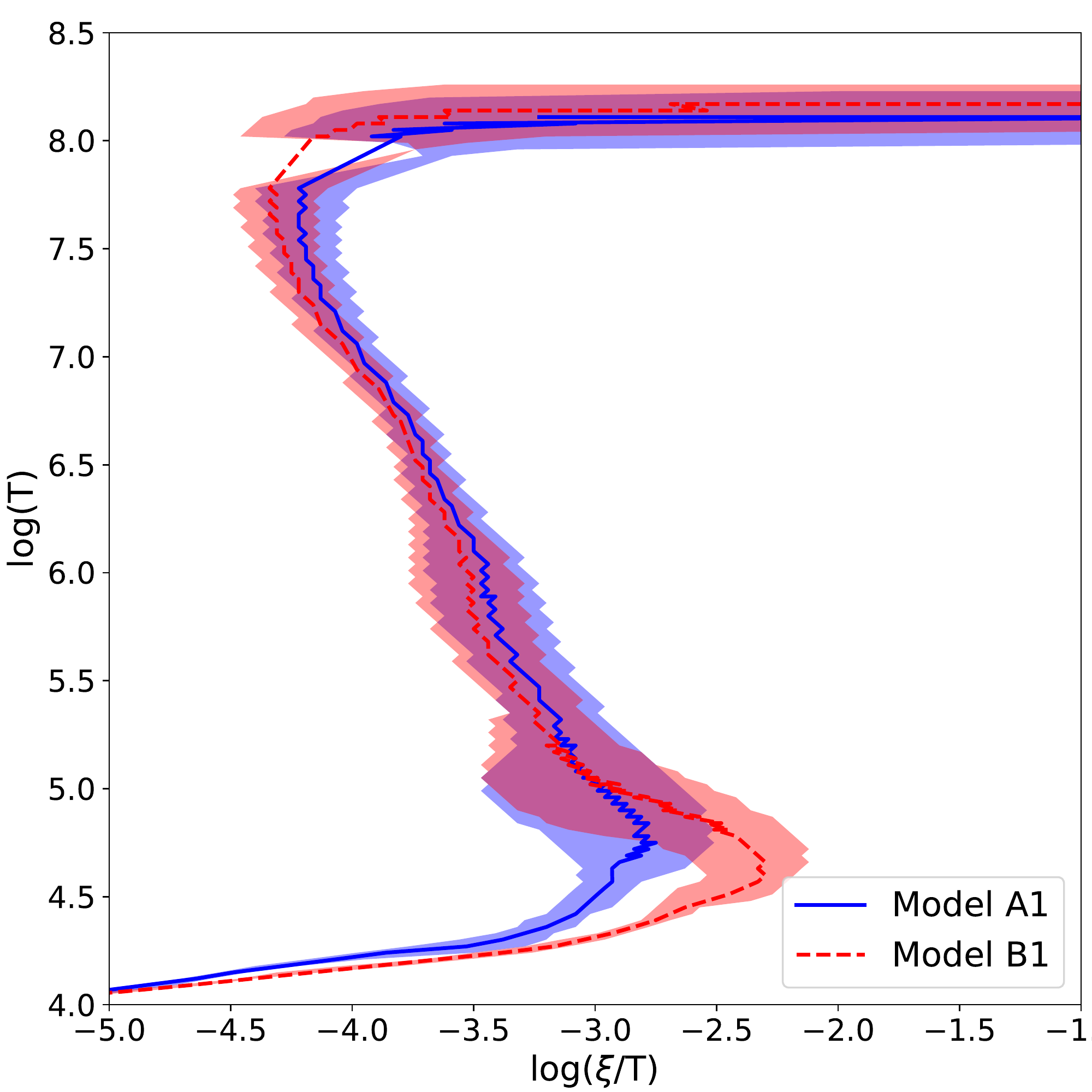} \\ 
         \includegraphics[scale=0.42]{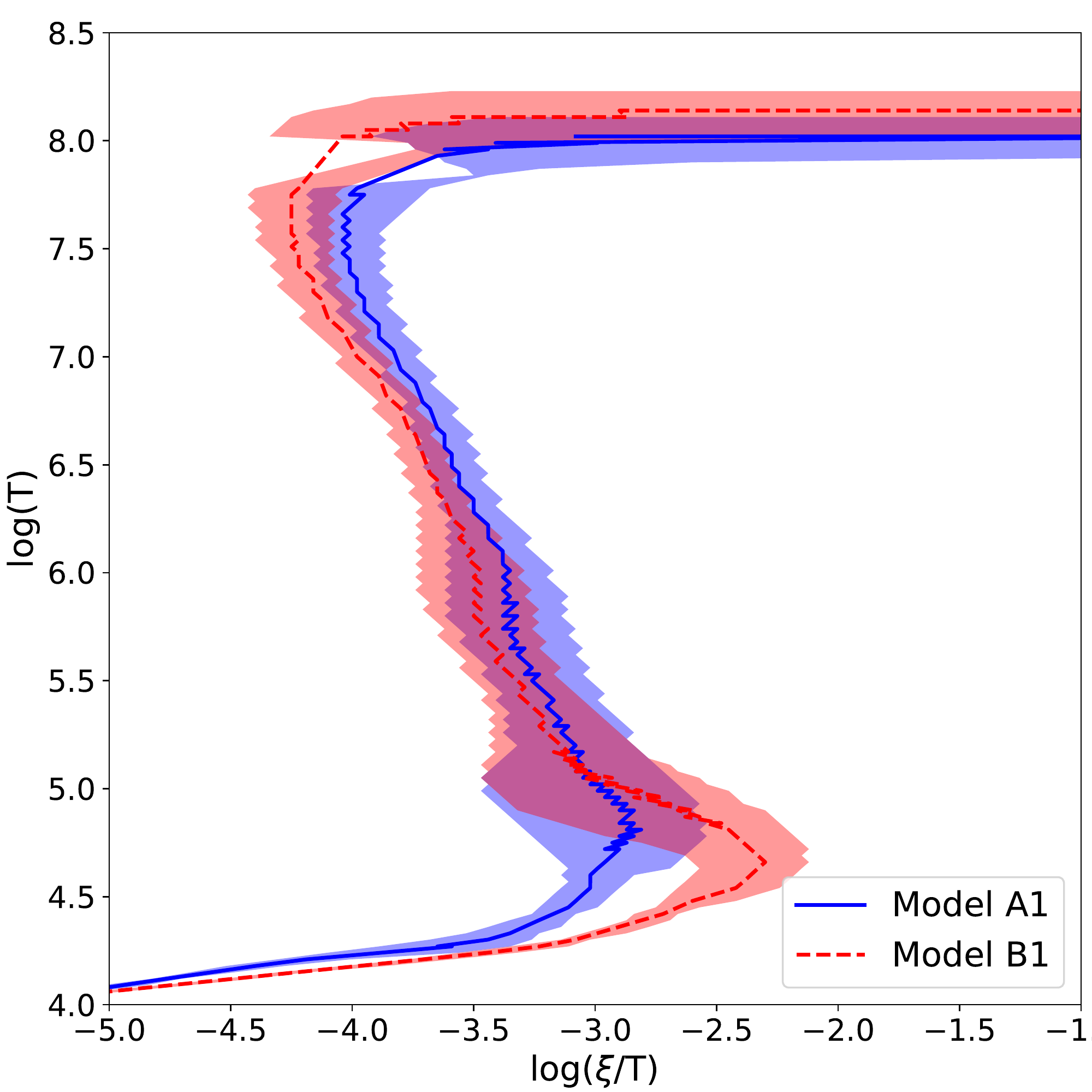}   
     \caption{Thermal stability curves obtained for Obs.~1 (top panel) and Obs.~2 (right panel).}\label{fig_stab_curves}
        \end{center}
   \end{figure}

   \begin{figure}
   \begin{center}
     \includegraphics[scale=0.42]{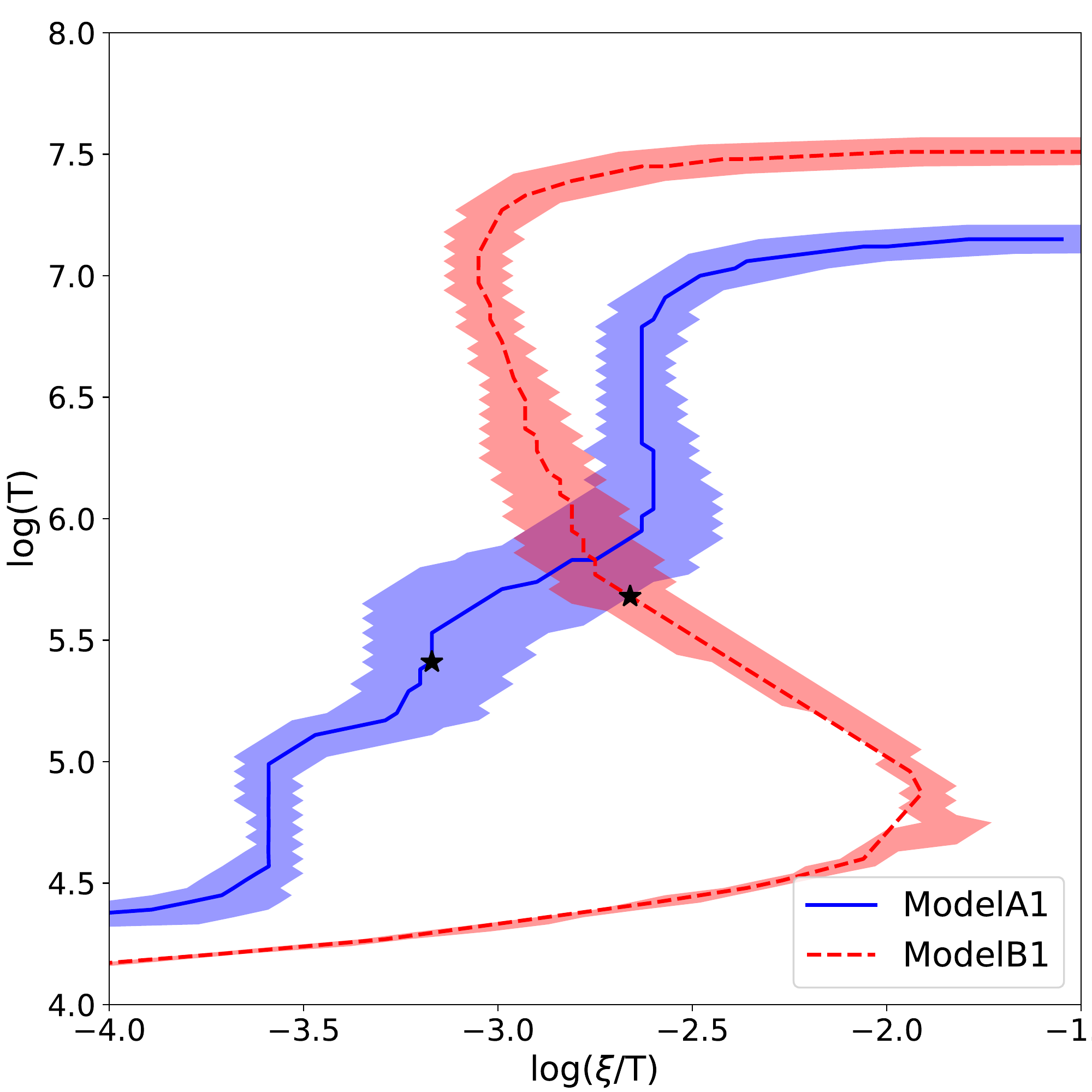}  
     \caption{Thermal stability curves obtained for Obs.~3. The black stars indicate the best-fit parameters obtained for Obs 3 with models A2 and B2.
     }\label{fig_stab_curves_obs3}
        \end{center}
   \end{figure}

\subsection{ISM multiphase X-ray absorption}\label{ism_3_obs} 
 
 In order to identify X-ray absorption features due to the ISM we substitute the {\tt tbabs} model with the {\tt IONeq} model, which assumes collisional ionization equilibrium and allows the modeling of absorption features due to the presence of multiple ISM phases, while the {\tt tbabs} model only includes the neutral component. In this way, and following \citet{gat18}, we used {\tt IONeq} to model the neutral (T$\approx 1\times 10^{4}$ K), warm (T$\approx 5\times 10^{4}$ K) and hot (T$\approx 2\times 10^{6}$ K) components of the ISM \citep{gat18}. It is important to model accurately these warm-hot ISM absorption features because they may be otherwise interpreted as local to the sources.  Following \citet{gat18}, we assumed $v_{turb}=75$ km s$^{-1}$ for the neutral-warm ISM component and $v_{turb}=60$ km s$^{-1}$ for the hot ISM component. The warm component is traced by ions such as {\rm Ne}~{\sc II}, {\rm Ne}~{\sc III}, {\rm Mg}~{\sc II} and {\rm Mg}~{\sc III} while the hot component is traced by ions such as {\rm Ne}~{\sc IX} and {\rm Mg}~{\sc XI}. In order to distinguish these models with those described in Section~\ref{model_first_case} we labeled them as model A2 and B2, respectively. 

First we modeled the EPIC-pn data and then we fitted the RGS spectra, by fixing the continuum parameters to those values obtained from the EPIC-pn fit (i.e. for the RGS fit, only the ISM column densities and the normalization are taken as free parameters). Best fit results are listed in Table~\ref{tab_con_gauss}. Figure~\ref{fig_warm_1_2_modeld} shows the residuals obtained for the EPIC-pn and RGS spectra using model B2. We note that the inclusion of the ISM multiphase component does not improve significantly the fits for Obs.~1 and Obs.~ 2 and that we only have found upper limits for the warm and hot ISM column densities, which are consistent between the different observations. For Obs.~3, the fit improves significantly when including the {\tt IONeq} component but residuals in the soft band are still apparent. Therefore, we proceed to included a photoionized absorber to improve the modeling of the soft X-rays energy band.

\subsection{Identification of a photoionized local absorber}\label{warm_obs3}
 
For the photoionized absorber, we used the {\tt warmabs} model, which is computed with the {\sc xstar} photoionization code \citep{kal01}. This code is designed to compute the physical conditions for an ionizing source surrounded by a gas taking into account physical processes such as  electron impact collisional ionization and excitation, radiative and dielectronic recombination, and photoionization. The main assumptions include a Maxwellian electron velocity distribution and ionization equilibrium conditions.  For each SED obtained from the continuum fitting described in Section~\ref{model_first_case}, we computed the energy level populations required by {\tt warmabs}.  

Figure~\ref{fig_warm_1_2_modeld} show the residuals of the fit for all three observations using model B3 (note that models A3 and B3 are similar in the XMM-newton energy band). Table~\ref{tab_ism} shows the best-fit parameters obtained for the EPIC-pn modeling when including the {\tt warmabs} component. We have fixed the $v_{turb}$ to 500 km/s. There is a significant improvement in the fits, from the statistical point of view, for Obs.~2 and Obs.~3 when including such a {\tt warmabs} component when modeling the EPIC-pn spectra (see the $\Delta\chi^{2}$ in Figure~\ref{fig_warm_1_2_modeld}).  However, when modeling the RGS observation, i.e. by fixing the continuum parameters to those obtained from the EPIC-pn fit and using the best-fit {\tt warmabs} parameters as initial values, we do not find such a significant improvement in the fit for Obs.~1 and 2.  It is possible that, depending of the brightness of the source, an intrinsic absorber can be identified only in the EPIC-pn spectra because of its larger effective area compared to the RGS instruments.  Given the similarity between both observations we did a test stacking their spectra and we have found that the RGS data also does not show a significant improvement when including the {\tt warmabs} component. The non-identification of such an absorber in the RGS spectra prevents us from drawing any firm conclusions from these data regarding the origin of the absorption for the first two observations. 

In the case of Obs.~3 we have found a significant improvement in the fits when including the {\tt warmabs} component in the model in both EPIC-pn and RGS spectra. In the case of EPIC-pn spectra we noted that the best-fit parameters (i.e. including the uncertainties) of the warm absorber are similar to those found for Obs.~1 and 2 (see Table~\ref{tab_ism}). The addition of the {\tt warmabs} component significantly improves the fit (e.g, from $\Delta \chi^{2}/d.o.f=464/122$ for Model~B2 to $\Delta \chi^{2}/d.o.f=352/120$ for Model~B3). Then, we fit the RGS spectra alone, by fixing the continuum parameters to those values obtained in the EPIC-pn fit and including a {\tt warmabs} component. In the case of {\tt warmabs}, we allow the $v_{turb}$ parameter to be free. The best-fit residuals for this case are shown in Figure~\ref{fig_warm_1_2_modeld}.  Table~\ref{tab_warm_3} shows the best-fit parameters obtained for models~A3 and~B3 for the RGS spectra. The {\tt constant} parameter accounts for the differences in normalization between RGS 1 and RGS 2. Once again, we found a notable improvement in the fits when including such a {\tt warmabs} component (e.g, $\Delta \chi^{2}/d.o.f=3055/1867$ for Model~B2 to $\Delta \chi^{2}/d.o.f=2948/1864$ for Model~B3). Finally, we have noticed that, if the continuum and ISM temperature parameters in the RGS analysis are free, a much better fit can be achieved ($\Delta \chi^{2}/d.o.f=2432/1857=1.31$ for Model~B3, see Figure~\ref{fig_best_fit}). However, given the small energy range covered, we prefer to fix the continuum parameters to those values obtained in the EPIC-pn fit. In the same way, if we set free the ISM temperatures and untie the $kT_{bb}$ from the {\tt nthcomp} and {\tt diskbb} components in the EPIC-pn analysis a much better fit can be achieved ($\Delta \chi^{2}/d.o.f=187/116=1.61$, see Figure~\ref{fig_best_fit}). In both cases the same {\tt warmabs} component still is required. We decided to perform the following analysis with the fits obtained from model A3 and B3.

\label{fig_best_fit}

Figure~\ref{fig_warm_3_flux} shows the best-fit obtained for the RGS spectra in counts (top panel) and flux (bottom panel) units, with labels indicating the main absorption lines identified in the spectra. Table~\ref{tab_warmabs_ions} shows the column densities derived from the {\tt warmabs} model for the most abundant ions for models A3 and B3. Obs.~3 shows an intrinsic ionized absorber ($1.96<\log\xi <2.05$) traced mainly by Ne\,{\sc x}, Mg\,{\sc xii}, Si\,{\sc xiii} and Fe\,{\sc xix}. We have found an upper limit for the outflow velocity of $<320$ km/s. The estimated location of the local static absorber obtained from the best-fit parameters varies in the range 9.5~$\times 10^{11}$ to 1.32~$\times 10^{12}$~cm (assuming a plasma density of $n_{e}=10^{12}$ cm$^{-3}$, as for the BH LMXB 4U~1630-47 in \citet{kub07}). We have performed a fit using Gaussians to estimate the equivalent widths of the main highly ionized lines identified in the RGS spectra of the Obs.~3 and the best-fit parameters obtained are listed in Table~\ref{tab_gauss_lines}. It is important to note that this local static X-ray absorber is observed simultaneously with a compact jet. 

To summarize, Obs.~1 and 2 reveal absorption at soft X-ray energies. However, without significant statistics in the RGS we cannot distinguish between absorption local to the source and in the ISM. Obs.~3, on the other hand, allows to determine that the there is a local photoionised component besides hot interstellar medium absorption and indicates that the absorption present in Obs.~1 and 2 has probably a similar origin.

   \section{Stability curves}\label{sec_seds}
    
The stability curve  (or thermal equilibrium curve) is a useful tool to study the equilibrium states of a photoionized plasma \citep{kro81}. It consist of a $T$ versus $\xi$/$T$ diagram which indicates, depending on the slope of the curve, the presence of a thermally stable region (positive slope) or a thermally unstable region (negative slope). Using the {\sc xstar} photoionization code (version~2.54) we compute stability curves to analyze the equilibrium conditions for the plasma associated to  IGR~J17091-3624. We ran a grid in the (log($T$),log($\xi$)) parameter space, with values ranging from $-4<$ log($\xi$) $<8$ (in units of erg cm s$^{-1}$) and $4<$ log($T$) <$10$ (in units of K).  We assumed an optically thin plasma with a constant density $n=10^{12}$ cm$^{-3}$ \citep[see][]{kub07} and solar abundances. For each (log($T$),log($\xi$)) point, the ionic fractions for all elements, as well as the heating and cooling rates, are stored. Then, we can determine those values corresponding to a thermal equilibrium state (i.e., heating=cooling).

Given that the stability curves are strongly affected by the shape of the SED \citep[see for example,]{kro81,cha09}, we compute stability curves for all best-fit models listed in Table~\ref{tab_ism}. Figure~\ref{fig_seds} shows the different SEDs obtained for the unabsorbed X-ray continuum.  Although the SEDs show differences at high energies, we note that without the inclusion of {\it Swift}/BAT data these differences would be even larger. The soft-energy region ($<$ 0.5~keV), on the other hand, is substantially different between the different models, due to the absence of data points in that energy range.

Figure~\ref{fig_stab_curves} shows the stability curves obtained for Obs.~1 and 2. The shaded region corresponds to the uncertainties of the stability curves for heating and cooling errors of 15 per cent. We note that, for Obs.~1 and ~2, the stability curves for the different models are similar for log($T$) $>5$, including a similar Compton temperature and the presence of few stable regions. Figure~\ref{fig_stab_curves_obs3} shows the stability curves obtained for Obs.~3 computed for both models considered. The best-fit parameters of the {\tt warmabs} component, described in Table~\ref{tab_warm_3}, lie in a thermally unstable region for model B1 while model A1 provides a thermally stable solution. These results shows the importance of the continuum modeling, ideally including a broad bandwidth coverage, in order to reach any conclusion about thermal instabilities that may cause the appearance of an absorber. 

\section{Discussion}\label{sec_dis}
 We do not find any signature of an UFO in our observations, as that previously reported for this source by \citet{kin12}.    We added Fe\,{\sc xxv} and Fe\,{\sc xxvi} lines with the same energy and width as those found by \citet{kin12} to our model and estimated upper limits of  $<$ 3.85$\times10^{-5}$ photons cm$^{-2}$ s$^{-1}$  and $<$ 4.66$\times10^{-5}$ photons cm$^{-2}$ s$^{-1}$for their normalisations, almost an order of magnitude below those reported by \citet{kin12}. Therefore, we can rule out that the absence of the UFO in our observations is due to lack of sensitivity. We note that our observations were taken in a different accretion state from that in which the UFO was detected, and so far only one observation from all available ones shows such phenomena \citep{jan15}.  

Instead, we have identified the presence of a local absorber in Obs.~3, during a hard-intermediate state. The lines do not show any significant blueshift down to 320 km/s. Our model includes both a collisional ionization equilibrium component (i.e. {\tt IONeq}, for the ISM) and a photoionization equilibrium component (i.e. {\tt warmabs}, for the local absorber). It is important to note that the RGS spectra allow us to distinguish between the two types of plasma. We found that a heartbeat pattern is not clearly distinguishable in the {\it XMM-Newton} light curves analyzed here (see Figure~\ref{fig_xmm_lc}). \citet{jan15} proposed that the presence of a wind may stabilize the disk and suppress the hearbeat pattern observed in the light curve and therefore an outflowing wind could be present in our observations. However, \citet{jan15} model depends on many parameters, including the ionization state and the velocities close to the black hole, which should be larger than the escape velocity. In this sense, outflowing wind may not always be detectable in spectroscopic observations.   

Low ionized state plasma, with $\log\xi\sim 1.8-2.5$,  has been identified in BH and NS LMXBs in the past \citep[see Table~1 in ][]{dia16}. However, we caution that none of these studies included in their models the warm-hot component associated to the ISM (but note that \citet{van09} included both a collisionally ionised and a photo-ionised plasma). Therefore, the column densities for the local absorber may be overestimated. When considering only the ISM, we obtain a high column density for the neutral component and upper limits for the warm and hot components. When including the {\tt warmabs} component, the fits improve by decreasing the column density of the ISM neutral component and increasing instead the warm-hot and {\tt warmabs} components to reduce the residuals.   

From all the sources with low ionisation plasma detections, at least EXO~0748--676 and MAXI~J1305--704 were in a hard accretion state when such plasmas were reported \citep{dia06,van09,shi13,mil14}, similarly to the observations reported in this paper. Different scenarios have been proposed to explain the presence of such ``atmospheric'' (i.e. bound to the source) low ionization plasma. \citet{van09} suggested that, depending on the geometry, such an absorber may trace the first segment of a circumbinary disc or the cold and compact region where the accretion stream hits the disc. Instead of a homogeneous distribution, \citet{shi13} suggested that the absorbers have clumpy and compact structures. If the absorber changes when the accretion state evolves from hard to soft state, then is possible that such plasma accelerates during the heating producing an outflowing wind. In the other case, it could be related to a more ``permanent'' structure whose detection would depend on source inclination.  Deep observations of BH LMXBs, taken during hard and soft accretion states, are crucial in order to address this issue.   

As was indicated in section \ref{warm_obs3}, a similar static absorber is identified in the first two observations but the statistic of the data (i.e. the number of counts for the RGS spectra) does not allow us to study its evolution during the outburst or its link, if any, with the existence of an outflowing wind. Finally, as was indicated in Section~\ref{sec_radio}, radio emission was detected simultaneously with the X-ray observations. In that sense, the local static X-ray absorber has been identified during a hard-intermediate accretion state simultaneously with a compact jet. While there are reported detections of jets simultaneously to winds \citep{lee02} or in similar states, even if not simultaneously \citep{hom16}, all of these detections are associated to states near or above the Eddington luminosity and the winds are highly ionised. In the cases of EXO~0748--676 and MAXI~J1305--704, there are no radio observations during the X-ray wind detections. Here, we observe for the first time a low ionization absorber simultaneously with a compact jet. Observations during the same state transition but at a later stage would help to clarify the potential interplay between this absorber, the jet, and a disc wind.

\section{Conclusions and summary}\label{sec_con}

 We have analyzed three {\it XMM-Newton} observations of the LMXB IGR~J17091-3624 that were taken during a transition from a hard accretion state to a hard-intermediate accretion state.  In one of the observations (Obs.~3), we have identified a local photoionised absorber with no significant blueshift in both EPIC-pn and RGS spectra which is traced mainly by Ne\,{\sc x}, Mg\,{\sc xii}, Si\,{\sc xiii} and Fe\,{\sc xviii}. This absorber may be present in the other two observations but is only detected at low significance due to the lower spectral statistics. This local static X-ray absorber is identified simultaneously with a compact jet. Such an absorber could be a permanent structure or a precursor of an outflowing wind. Future X-ray observations of bright LMXBs, with high inclination and during hard accretion states, will help to better understand the origin of such plasma.

 \section{Acknowledgements}
We thank the anonymous referee for the careful reading of our manuscript and the valuable comments.. The Australia Telescope Compact Array is part of the Australia Telescope National Facility which is funded by the Australian Government for operation as a National Facility managed by CSIRO. JCAM-J is the recipient of an Australian Research Council Future Fellowship (FT140101082), funded by the Australian government.

\bibliographystyle{mnras}

\end{document}